\documentclass[bibyear]{aa}
\usepackage{deluxetable}
\usepackage{xtab,booktabs}
\usepackage{caption}
\usepackage{lineno}

\usepackage{natbib}
\bibpunct{(}{)}{;}{a}{}{,} 

\usepackage[colorlinks=true, linkcolor=blue, citecolor=blue, filecolor=blue, urlcolor=blue]{hyperref}
\newcommand{\nraoblurb}{The National Radio Astronomy Observatory is
a facility of the National Science Foundation operated under cooperative
agreement by Associated Universities, Inc.}
\newcommand{\hide}[1]{}

\newcommand{\lsim}{\ensuremath{\,\lesssim\,}\xspace}
\newcommand{\micron}{\mbox{$\mu$m}}%
\newcommand{\gl}{\ensuremath{\ell}\xspace}
\newcommand{\gb}{\ensuremath{{\it b}}\xspace}
\newcommand{\absb}{\ensuremath{\vert\,\gb\,\vert}\xspace}

\newcommand{\lb}{\ensuremath{(\gl,\gb)}\xspace}

\newcommand{\mhz}{\ensuremath{\,{\rm MHz}}\xspace}
\newcommand{\ghz}{\ensuremath{\,{\rm GHz}}\xspace}

\newcommand{\degree}{\ensuremath{^\circ}\xspace}

\newcommand{\hi}{{\rm H\,{\footnotesize I}}\xspace}
\newcommand{\hii}{{\rm H\,{\footnotesize II}}\xspace}

\begin{document}

\title{Supernova Remnant Candidates Discovered by the SARAO
MeerKAT Galactic Plane Survey}

\author{L.~D.~Anderson\inst{\ref{wvu},}\inst{\ref{gbo},}\inst{\ref{center}}
\and F.~Camilo\inst{\ref{sarao}}
\and Timothy Faerber\inst{\ref{wvu},}\inst{\ref{center}}
\and M.~Bietenholz\inst{\ref{hartebeesthoek},}\inst{\ref{toronto}}
\and C.~Bordiu\inst{\ref{inaf}}
\and F.~Bufano\inst{\ref{inaf}}
\and J.~O.~Chibueze\inst{\ref{j1},}\inst{\ref{j2},}\inst{\ref{j3}}
\and W.~D.~Cotton\inst{\ref{sarao},}\inst{\ref{nrao}} 
\and A.~Ingallinera\inst{\ref{inaf}}
\and S.~Loru\inst{\ref{inaf}}
\and A.~Rigby\inst{\ref{leeds}}
\and S.~Riggi\inst{\ref{inaf}}
\and M.~A.~Thompson\inst{\ref{leeds}}
\and C.~Trigilio\inst{\ref{inaf}}
\and G.~Umana\inst{\ref{inaf}}
\and G.~M.~Williams\inst{\ref{leeds},}\inst{\ref{aberystwyth}}
}

\institute{Department of Physics and Astronomy, West Virginia University, Morgantown WV 26506, USA\label{wvu}
\and Adjunct Astronomer at the Green Bank Observatory, P.O. Box 2, Green Bank WV 24944, USA\label{gbo}
\and Center for Gravitational Waves and Cosmology, West Virginia University, Chestnut Ridge Research Building, Morgantown, WV 26505, USA \label{center}
\and South African Radio Astronomy Observatory, 2 Fir Street, Observatory 7925, South Africa\label{sarao}
\and SARAO/Hartebeesthoek Radio Astronomy Observatory, PO Box 443, Krugersdorp 1740, South Africa\label{hartebeesthoek}
\and Department of Physics and Astronomy, York University, Toronto, M3J 1P3, Ontario, Canada\label{toronto}
\and INAF -- Osservatorio Astrofisico di Catania, Via S. Sofia 78, I-95123 Catania, Italy\label{inaf}
\and Department of Mathematical Sciences, University of South Africa, Cnr Christian de Wet Rd and Pioneer Avenue, Florida Park, 1709, Roodepoort, South Africa\label{j1}
\and Centre for Space Research, Physics Department, North-West University, Potchefstroom 2520, South Africa\label{j2}
\and Department of Physics and Astronomy, Faculty of Physical Sciences, University of Nigeria, Carver Building, 1 University Road, Nsukka 410001, Nigeria\label{j3}
\and National Radio Astronomy Observatory, 520 Edgemont Road, Charlottesville, VA 22903, USA\label{nrao}
\and School of Physics and Astronomy, University of Leeds, Leeds LS2 9JT, UK\label{leeds}
\and Department of Physics, Aberystwyth University, Ceredigion, Cymru, SY23 3BZ, UK\label{aberystwyth}
}


\abstract
{Sensitive radio continuum data could remove the difference between the number of known supernova remnants (SNRs) in the Galaxy compared to that expected, but due to confusion in the Galactic plane, faint SNRs can be challenging to distinguish from brighter \hii\ regions and filamentary radio emission.} {We wish to exploit new SARAO MeerKAT 1.3\,\ghz\ Galactic Plane Survey (SMGPS) radio continuum data, which covers $251\degree \le \ell \le  358\degree$ and $2\degree \le \ell \le  61\degree$ at $\absb \le 1.5\degree$, to search for SNR candidates in the Milky Way disk.} {We also use MIR data from the Spitzer GLIMPSE, Spitzer MIPSGAL, and WISE surveys to help identify SNR candidates. The identified SNR candidate are sources of extended radio continuum emission that lack MIR counterparts, are not known as \hii\ regions in the WISE Catalog of Galactic \hii\ Regions, and are not known previously as SNRs} {We locate 237
new Galactic SNR candidates in the SMGPS data.  We also identify and confirm the expected radio morphology for 201
objects listed in the literature as being SNRs and 130
previously-identified SNR candidates.  The known and candidate SNRs have similar spatial distributions and angular sizes.  
} {The SMGPS data allowed us to identify a large population of SNR candidates that can be confirmed as true SNRs using radio polarization measurements or by deriving radio spectral indices.  If the 
237 candidates are confirmed as true SNRs, it would approximately double the number of known Galactic SNRs in the survey area, alleviating much of the difference between the known and expected populations.}

\titlerunning{MEERKAT SNRs}

\maketitle

\section{Introduction}
\label{sec:intro}
The number of confirmed Galactic SNRs numbers only a few hundred according to the two most authoritative catalogs: that of \citet[][hereafter ``G22'']{green22}\footnote{https://www.mrao.cam.ac.uk/surveys/snrs/}, which contains 303 SNRs\footnote{The most recent published version is that of \citet{green14a}; we use here the version found online.}, and ``SNRCat'' \citep{ferrand12}, which lists 383 objects in the online catalog\footnote{http://snrcat.physics.umanitoba.ca/}.  Based on OB star counts, pulsar birth rates, Fe abundances, and the SN rate in other Local Group galaxies, there should be $\ga 1000$ Galactic SNRs \citep{li91, tammann94}. \citet{ranasinghe22} suggested that there actually may be a few thousand Galactic SNRs based on the population detected to date and on SNR search selection effects.  The difference between the identified population of Galactic SNRs and the expected number is likely due to a lack of sensitive radio continuum data and confusion
in the Galactic plane \citep[e.g.,][]{brogan06, ranasinghe22}.

The Galactic supernova rate is key to understanding the properties and dynamics of the Milky Way.  The number of SNRs in the Galaxy is related to recent massive star formation activity \citep[cf.][]{tammann94}.  Supernovae inject energy into the interstellar medium (ISM), driving molecular cloud turbulence and galactic fountains out of the disk \citep{deavillez05, joung09, padoan16, girichidis16};  they therefore affect the disk scale height and star formation properties of a galaxy \citep{ostriker10,  ostriker11, faucher13}. 

SNRs can be most efficiently identified using radio data.  According to our assessment of the G22 catalog, $\sim\!90$\% of known SNRs are detected and well-defined in the radio regime, $\sim\!40\%$ detected in X-rays, and $\sim\!30\%$ in the optical. More sensitive radio observations may increase the number detected further.  The most common radio morphology in the G22 catalog, $\sim\!80\%$ of SNRs, is that of a limb-brightened shell or a partial shell, where the diameter of the shell is set by  the expanding shock wave produced by the explosion.  SNRs can also be classified as ``filled-centre'' or ``plerions,'' if they have centrally concentrated radio emission, which is is generally caused by a pulsar wind nebula \citep[e.g., the Crab nebula,][]{weiler88, bietenholz15b}, ``composite,'' for which the SNR appears to have both a shell and an internal component (either a nonthermally-emitting pulsar-driven nebula or a thermally-emitting x-ray source)
\citep[see][for a review of radio SNR morphologies]{dubner15}.


Separating nonthermally-emitting SNRs from the much more common thermally emitting \hii\ regions in the Galactic plane is challenging.  There are $\gtrsim\!7000$ identified Galactic \hii\ regions \citep{armentrout21}.
If there are data at multiple radio frequencies, one can compute the spectral index $\alpha$, where $S\propto 
\nu^\alpha$ with $S$ being the flux density and $\nu$ the frequency\footnote{We are using the convention that positive values of $\alpha$ indicate increasing flux densities with frequency.}.  Most SNRs have spectral indices in the range from $\alpha=-0.8$ to $\alpha = 0.0$, whereas \hii\ regions should have $\alpha \sim\!-0.1$ if optically thin and $\alpha > -0.1$ if optically thick.  Although a spectral index value near 0 does not discriminate between the two classifications, values $\alpha\lsim-0.2$ indicate nonthermal emission, and therefore an SNR rather than an \hii\ region.
The computation of the spectral index, however, is difficult because all radio continuum datasets must have sensitivity to the same spatial scales.  Also, filled-center SNRs may have spectral indices near $-0.1$ \citep[e.g.,][find $\alpha = 0.08$ for G21.5$-$0.5]{bietenholz08}.
An additional indication of nonthermal emission may come from the presence of polarized radio continuum emission, although due to the fact that the polarized emission fraction is often quite low, this criterion is also difficult to apply \citep{dokara18}.  In a SNR, the intrinsic polarization angle usually follows the bright ridges of emission \citep[e.g.,][]{cotton24}, but the presence of polarized emission in the same direction as a SNR does not necessarily mean that it is a SNR. 

An easier and perhaps more reliable criterion for separating \hii\ regions from SNRs is the mid-infrared (MIR) to radio continuum flux ratio.  Although MIR emission can be detected from SNRs in
some cases \citep{reach06, pinheiro11}, many
researchers have shown that SNRs are deficient in MIR emission
compared to \hii\ regions \citep[e.g.,][]{cohen01, pinheiro11}.  The MIR to radio flux ratio for SNRs is about 100 times lower than that of optically thin \hii\ regions.  

In the last decade, studies of radio continuum and infrared data have identified a large number of Galactic supernova remnant (SNR) candidates.  Those that have been confirmed as SNRs are in the SNR catalogs discussed previously, but the majority are awaiting confirmation. 
 \citet{green14b} used the anti-correlation between radio and 8\,\micron\ emission to locate 23 new SNR candidates from 843\,\mhz\ Molonglo Galactic Plane Survey (MGPS) data.  \citet{anderson17} used 1.4\,\ghz\ continuum data from The \hi, OH, Recombination line survey of the Milky Way \citep[THOR;][]{beuther16}, combined with the 1.4\,\ghz\ radio continuum data from the VLA Galactic Plane Survey \citep[VGPS][]{stil06}, to identify 76 new SNR candidates.  \citet{dokara18} confirmed two of these 76 candidates as true SNRs using spectral index measurements.  \citet{hurley-walker19} found 27 new SNRs using data from the GaLactic and Extragalactic All-sky Murchison Widefield Array (GLEAM) survey and confirmed 26 of them using spectral index measurements; these 26 are listed as known SNRs in G22. 
\citet{dokara21} used data from the $4-8$\,\ghz\ GLOSTAR survey to identify 157 SNR candidates, 
9 of which have nonthermal emission based on polarization measurements. 
And, finally, \citet{ball23} identified 13 new Galactic SNR candidates in 933\,\mhz\ Australian Square Kilometre Array Pathfinder (ASKAP) data.  These studies followed those of \citet{helfand06}, which discovered 49 new SNR candidates in The Multi-Array Galactic Plane Imaging Survey (MAGPIS) 20\,cm data, and \citet{brogan06}, which discovered 35 SNR candidates in their VLA data.  Even if all the recently-identified SNR candidates are confirmed as true SNRs, there remains a difference between the number of detected Galactic SNRs compared to that expected.

Here, we use new 1.3\,\ghz\ SARAO
MeerKAT Galactic Plane Survey data (``SMGPS'') to identify SNR candidates in the inner Galaxy.  This work complements that of Bordiu et al. (2024, submitted), who used SMGPS data to create an extended source catalog, and Loru et al. (2024, submitted), who provide an in-depth study of 28 Galactic SNRs in SMGPS data for which flux and spectral indices can be derived.

\section{Data}
\subsection{SARAO
MeerKAT Galactic Plane Survey (SMGPS)}
The ideal radio continuum dataset for SNR searches is sensitive to large, extended structures but also boasts high angular resolution.  Extended source sensitivity is necessary for the detection of low surface brightness SNRs and high angular resolution allows one to disentangle the complicated emission of the Galactic plane.

The 1.3\,\ghz\ SMGPS \citep{goedhart24}\footnote{Data can be found here: \url{https://doi.org/10.48479/3wfd-e270}. When using DR1 products,
\citet{goedhart24} should be cited, and the MeerKAT telescope acknowledgement
included.} covers 
$251\degree \le \ell \le  358\degree$ and $2\degree \le \ell \le  61\degree$ at $\absb \le 1.5\degree$.  The boundaries at high and low Galactic latitudes are slightly irregular because the survey follows the Galactic warp. 
 A full description of the radio observations and the data reduction
procedures is provided in \citet{goedhart24}.  Observations used the 64 antenna MeerKAT array in the Northern Cape Province of South Africa, which is described in \citet{Jonas2016}, \citet{Camilo2018}, and \citet{DEEP2}.  We use here the ``zeroth moment'' data, which have a central frequency of 1293\,\mhz with a total used bandwidth of 672\,\mhz \citep[these values both decrease at high and low Galactic latitudes; see][]{goedhart24}.
The angular resolution is $\sim\!8\arcsec$ and the data are sensitive to emission up to angular scales of $\sim\!30\arcmin$.  The background rms noise in areas far from the Galactic Plane and bright point sources is $\sim\!30\mu{\rm Jy\,beam}^{-1}$.  The low surface brightness noise threshold together with the
sensitivity to both large and small-scale structures makes the SMGPS survey the ideal
data set to identify new SNRs.

\subsection{MIR data}
Over the zone $65\degree > \ell > -100\degree$, $\absb < 1.0\degree$ we use Spitzer 8.0 and 3.6\,\micron\ data from the  GLIMPSE survey \citep{benjamin03, churchwell09} and 24\,\micron\ data from the MIPSGAL survey \citep{carey09}.  Outside this zone we use data from the all-sky Widefield Infrared Survey Explorer \citep[WISE;][]{wright10} at 12 and 22\,\micron.  Since the GLIMPSE and MIPSGAL surveys have better angular resolution and sensitivity than that of WISE, we use Spitzer data when possible.

\hii\ regions have strong emission at these wavelengths. The $\sim\!10\,\micron$ emission is largely from polycyclic aromatic hydrocarbons (PAHs), which fluoresce in the presence of soft ultra-violet ($\sim\!5\,$eV) radiation \citep{voit92} and the $\sim\!20\,\micron$ emission is mainly from small dust grains cospatial with the \hii\ region plasma.  

For most SNRs, MIR emission at these wavelengths is largely absent, or at least is at a low intensity compared to that of \hii\ regions \citep[e.g.,][]{cohen01, pinheiro11}.  Some young SNRs do have strong MIR emission, however, especially at $\sim\!20\,\micron$.  The origin of this emission is dust, atomic/molecular line emission, or synchrotron emission, with the relative importance of each depending on the SNR in question \citep[see][for a concise summary]{goncalves11}.

\subsection{The WISE Catalog of Galactic HII Regions\label{sec:wise}}
The WISE Catalog of Galactic \hii\ Regions \citep[][hereafter the ``WISE Catalog'']{anderson14} is the largest catalog of Galactic \hii\ regions.  All catalog entries have WISE \citep{wright10}  $\sim\!20$\,\micron\ emission surrounded by $\sim\!10$\,\micron\ emission \citep[][]{anderson11}.  
The WISE catalog contains $\sim\!8000$ objects with the MIR morphology of \hii\ regions.  The $\sim\!20$\,\micron\ emission from \hii\ regions is caused by small stochastically heated dust grains that are mixed with the \hii\ region plasma, while the $\sim\!10$\,\micron\ intensity is dominated by emission from PAHs.  All known Galactic \hii\ regions have this characteristic morphology.  
Here, we use Version~3.0 of the catalog\footnote{http://astro.phys.wvu.edu/wise}.

\subsection{SNR Catalogs\label{sec:known_cat}}
The G22 catalog currently contains 303 regions compiled from the literature.  \citet{green04} suggest that a previous (but similar) version of the catalog was largely complete to a radio surface density limit of $10^{-20}$\,W\,m$^{-2}$\,Hz$^{-1}$\,sr$^{-1}$ (1\,MJy\,sr$^{-1}$).  At the $8\arcsec$ resolution of the SMGPS data, this corresponds to 1.7\,mJy\,beam$^{-1}$, or $\sim\!60$ times greater than the SMGPS $1\sigma$ point source sensitivity.  In addition to the surface brightness limit, the catalog appears to be lacking the small angular size SNRs that are expected \citep{green15}.

SNRCat \citep{ferrand12} contains 383 entries in the most recent online version and focuses largely on their high energy emission.  All G22 sources are also included in SNRCat.  Even though there are more entries than in G22, many of these are not the same as the SNRs listed in G22; SNRCat includes pulsar wind nebulae, bow-shock nebulae, high-energy (X-ray and
gamma-ray) discovered SNRs, and magnetar-hosting SNRs.

As there are significant differences in their compositions, we treat G22 and SNRCat separately.  Although both catalogs cover the entire sky, since they are not derived from homogeneous data sets, the catalog sensitivities vary with Galactic location.  
Both catalogs contain at a minimum spatial coordinates and angular sizes for all entries.  When the extent of the SNR is listed with an ellipse, we instead use a circle that has an angular radius that is an average between that of the semimajor and semiminor axes.

\subsection{SNR Candidate Catalogs\label{sec:cand_cat}}
There are numerous extant catalogs of SNR candidates.  
We primarily use the compilation of previously-known and newly-discovered SNR candidates in \citet{dokara21}, which includes results from \citet{brogan06,helfand06, anderson17, hurley-walker19}.  We supplement this list using the results from numerous other studies compiled in the G22 documentation and additionally add the recent study of
\citet{ball23}.
Over the SMGPS range, there are 190SNR candidates; 74 of these were compiled by \citet{dokara21}, and 78 were identified for the first time by \citet{dokara21}.  

As with the known SNRs, in cases where the extent of the SNR candidate is defined with an ellipse, we instead use a circle that has a radius in between that of the semimajor and semiminor axes.

\section{Method}
The present study has two main goals: 1) to assess the radio emission from known and candidate SNRs in the SMGPS data and 2) to identify new SNR candidates.

For the first goal, we inspect SMGPS and MIR data at the positions of all known and candidate SNRs from the catalogs in Sections~\ref{sec:known_cat} and \ref{sec:cand_cat}.  We take the coordinates of a given object from the most recent published study, which is not necessarily the same as that listed in the catalogs.  During this visual inspection, we primarily seek to ascertain if the identified SNR or SNR candidate is detected in the SMGPS.  If it is, we determine if it is confused with an \hii\ region.

For the second goal, our identification methodology is similar to that used by
numerous recent authors \citep[e.g.,][]{anderson17, dokara21, ball23}: we identify discrete regions of radio continuum
emission that 1) have a radio morphology consistent with known SNRs (i.e. roughly circular) , 2)~are not identified as being \hii\ regions in the WISE Catalog, known SNRs, or SNR candidates, and 3)~lack
  Spitzer or WISE MIR emission characteristic of thermal sources.
Criteria~\#2 and \#3 are somewhat redundant, as nearly all
discrete sources of coincident MIR and radio continuum emission in the
Galactic plane are \hii\ regions and are included in the WISE
catalog.  As mentioned previously, some SNRs do have MIR emission, but the quality of this emission differs between SNRs and \hii\ regions.  Aside from the radio-to-MIR flux ratio being higher for SNRs than for \hii\ regions \citep[e.g.,][]{cohen01, pinheiro11}, the $\sim\!24\,\micron$ emission for SNRs more closely follows the radio than for \hii\ regions, and there is no related $\sim\!10\,\micron$ emission.  

\begin{figure*}
    \centering
    \includegraphics[width=7.0in]{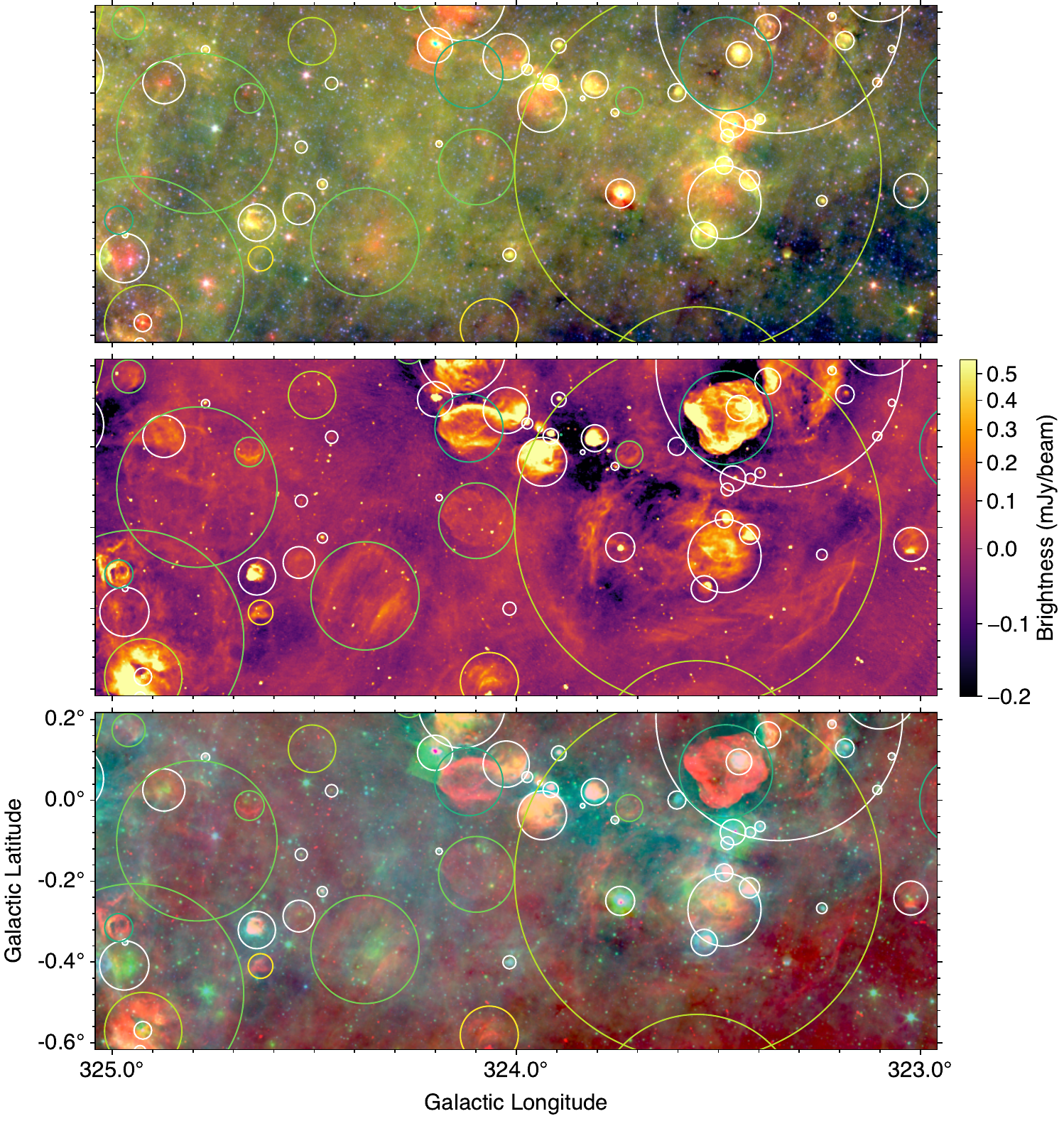}
    \caption{Example field centered at \lb = (324.0\degree, $-$0.2\degree).  The top panel shows Spitzer three-color data, with MIPSGAL 24\,\micron\ data in red, GLIMPSE 8.0\,\micron\ data in green, and GLIMPSE 3.6\,\micron\ data in blue.  The middle panel shows the 1.3\,\ghz\ SMGPS images.  The bottom panel has 1.3\,\ghz\ SMGPS images in red, MIPSGAL 24\,\micron\ data in green and GLIMPSE 8.0\,\micron\ data in blue.
    White circles show \hii\ regions from the WISE Catalog (Section~\ref{sec:wise}), blue-green circles show previously-known SNRs (Section~\ref{sec:known}), green circles show  previously-known SNR candidates (Section~\ref{sec:known_cand}), light green circles show SNR candidates newly identified here (Section~\ref{sec:cand}), and yellow circles show ``Unusual'' sources (Section~\ref{sec:odd}). 
 Although there are no examples in this field, in subsequent figures we show misidentified or nondetected SNRs and SNR candidates with dashed circles.\\\\\\\\\\\\\\\\\\\\\\\\\\}\label{fig:overview}
\end{figure*}

We illustrate the by-eye identification process in Figure~\ref{fig:overview}. We first identify the SMGPS emission associated with all known SNRs and SNR candidates (red and green circles, respectively).  We then identify all discrete, extended radio continuum sources that are not associated with these SNRs or SNR candidates, and also are not associated with \hii\ regions in the WISE Catalog.  We classify such sources as new SNR candidates (light green) or unusual objects (pink).  We inspect each SMGPS field on three separate occasions to ensure that we identify as many SNR candidates as possible. 

We do not have a preferred morphology for the identified SNR candidates aside from 
requiring that the emission is relatively circularly symmetric.  We then search the MIR data for emission of a complementary morphology to that of the SMGPS data, which allows us to remove any remaining \hii\ regions not included in the WISE Catalog.  We classify the remaining radio continuum sources are SNR candidates.

We approximate the center and angular extent of each identified object by defining a circle that encloses its SMGPS emission.  Our method therefore defines new positions for all known and candidate SNRs in the SMGPS zone.  For partial shells, the defined circle follows the curvature of the shell.  Too, if the source is partially off the SMGPS zone, we define the circle following the emission that is detected; such centroids and radii can therefore be quite uncertain.

For each newly-identified SNR candidate, we query the Simbad database\footnote{https://simbad.u-strasbg.fr/simbad/}, with a search radius of $5\arcmin$.  This process allows us to remove previously-identified objects from the sample.

\section{Results}

We summarize the results of our investigation of previously-known SNRs, previously-identified SNR candidates, and newly-discovered sources in Table~\ref{tab:summary}.  This table lists for each category the number of objects, and the number confirmed as SNRs or SNR candidates (if previously-known).  We give more details on these samples in the following subsections.  

Images and SMGPS FITS cutouts for all objects studied are hosted on \url{https://doi.org/10.48479/0n8c-5q84}.

\begin{table}
 \caption{Summary of Sources Studied \label{tab:summary}}
 \begin{tabular}{lrr}

 \hline\hline 
 Category & Number & Confirmed\\\hline
     Known SNRs & 238 & 201 \\
     ~~~~G22 & 187 & 184\\
     ~~~~SNRCat (unique)\tablefootmark{a} & 51 & 17\\
     Known SNR Candidates & 170 & 130\\
     New SNR Candidates\tablefootmark{b} & 237 & \nodata\\
     ~~~~Class I & 83 &\nodata\\
     ~~~~Class II & 104&\nodata\\
     ~~~~Class III & 50&\nodata\\
     Unusual   & 49 & \nodata\\\hline
\end{tabular}
\tablefoot{\tablefoottext{a}{``Unique'' refers to sources only in SNRCat, but not in G22.} \tablefoottext{b}{New SNR candidates are classified from I to III in order of decreasing reliability; see Section~\ref{sec:cand}.}}
\end{table}
     
\subsection{Previously-Identified SNRs\label{sec:known}}
\begin{figure*}
    \centering
    \includegraphics[width=3.6in]{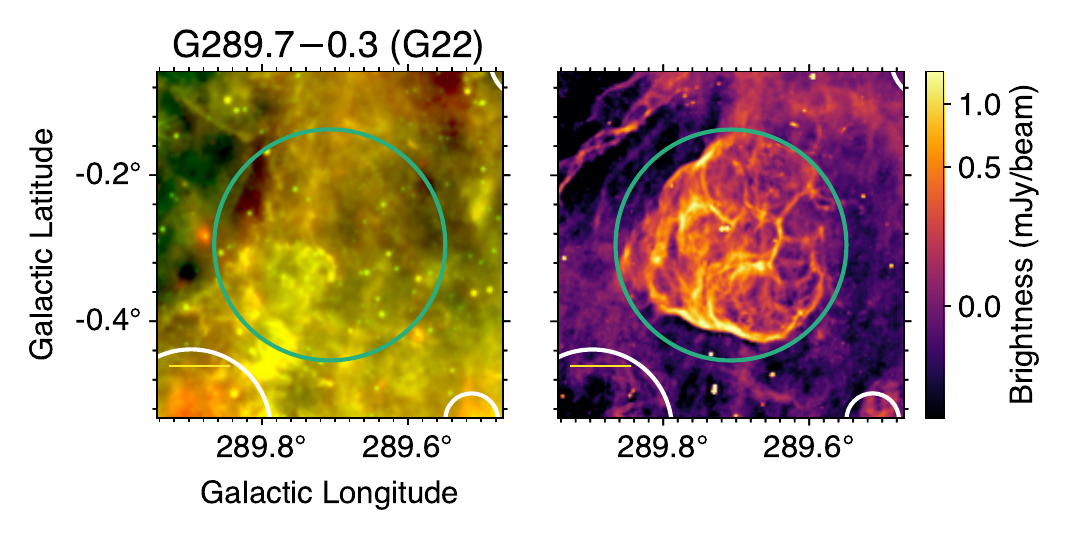}
    \includegraphics[width=3.6in]{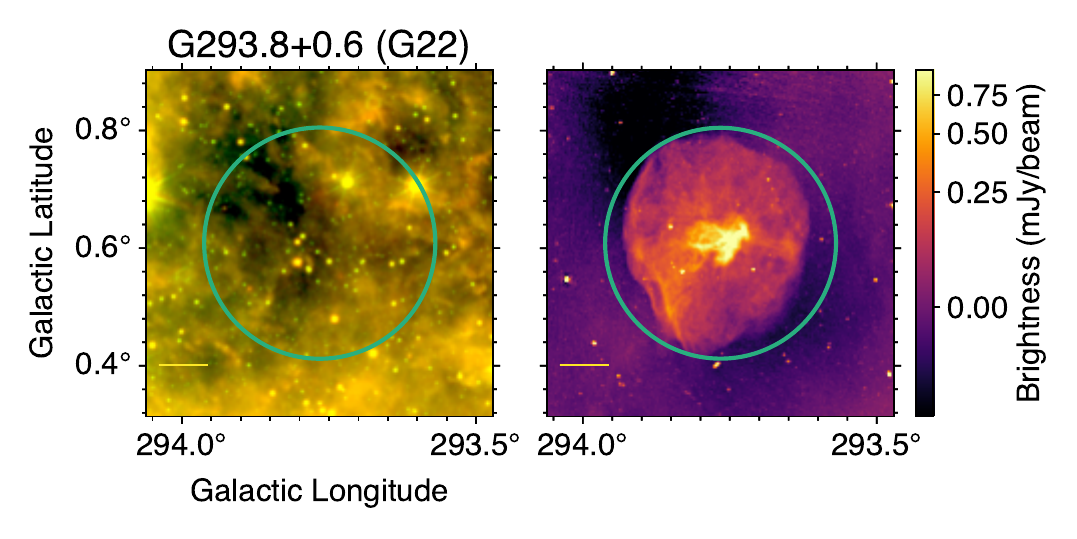}
    \includegraphics[width=3.6in]{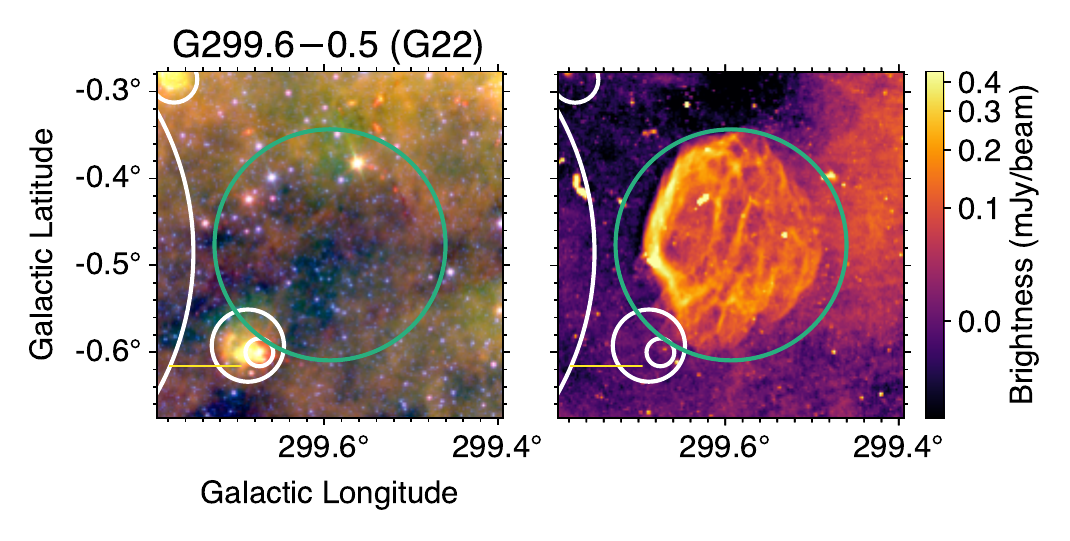}
    \includegraphics[width=3.6in]{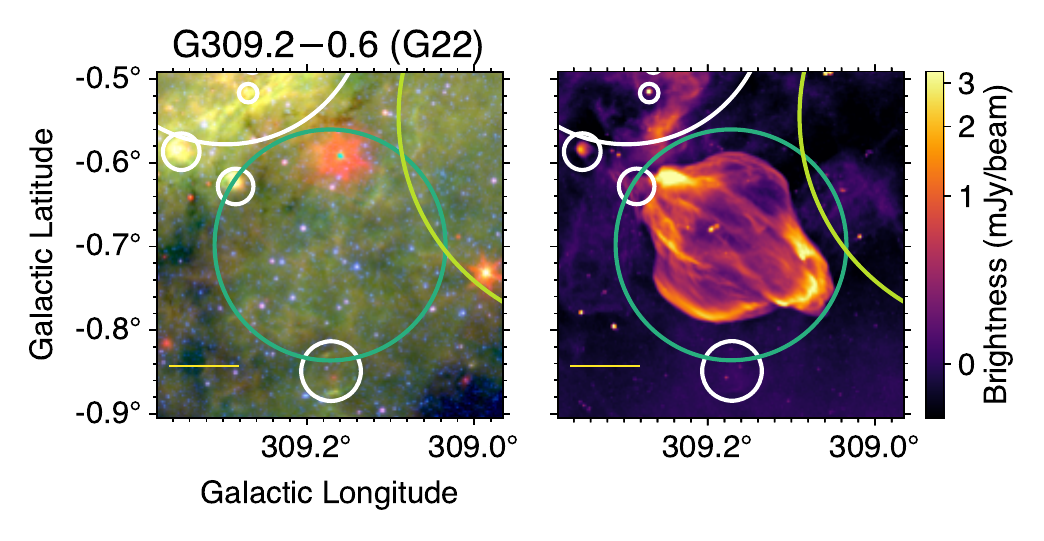}
    \includegraphics[width=3.6in]{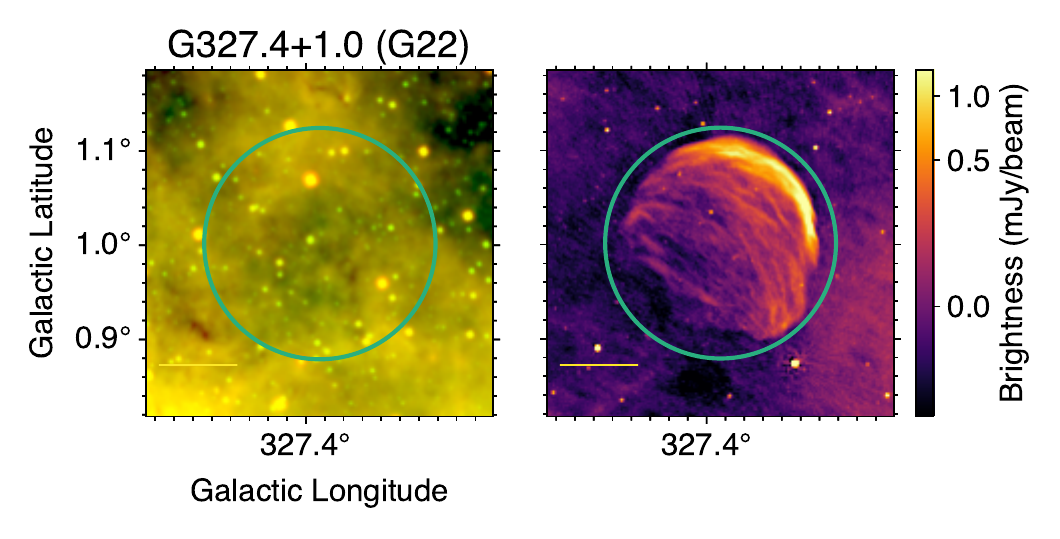}
    \includegraphics[width=3.6in]{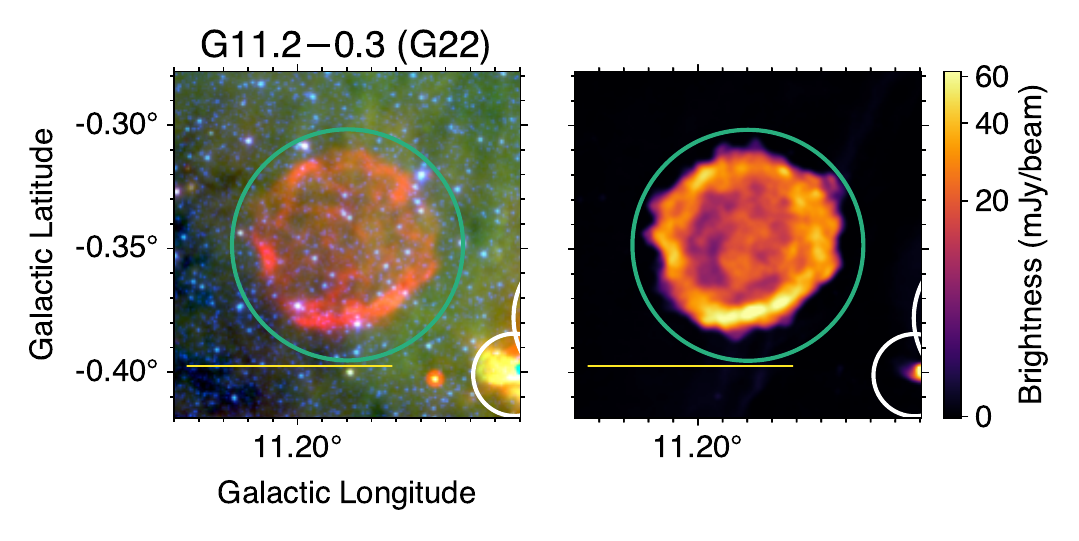}
    \includegraphics[width=3.6in]{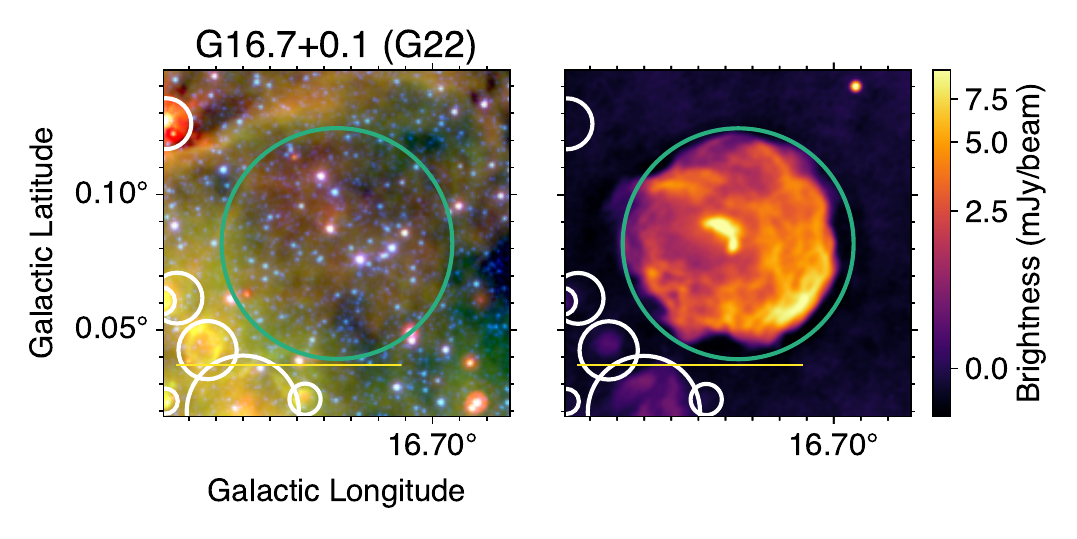}
    \includegraphics[width=3.6in]{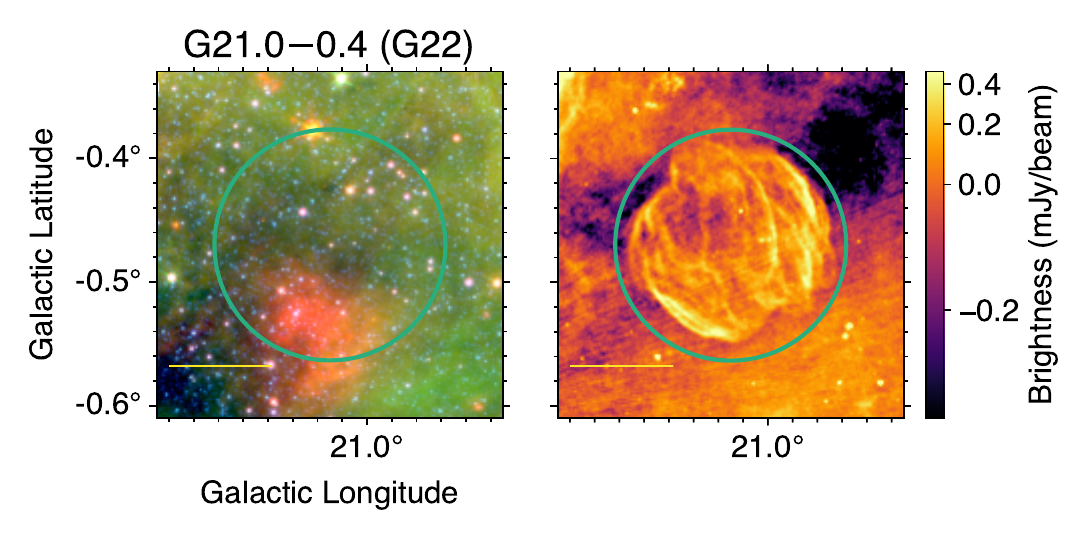}
    \caption{Objects from the G22 catalog.  MIR data is in the left panels (Spitzer or WISE) and SMGPS 1.3\,\ghz\ data are on the right.  The symbols have the same meaning as in Figure~\ref{fig:overview}, with the G22 SNRs being represented by blue-green circles at the centers of all panels.  Cyan scale bars at the lower left of each panel are $5\arcmin$ long.}
    \label{fig:known_G22}
\end{figure*}

We detect SMGPS emission from all
187 G22 SNRs whose centroids fall within the SMGPS area.  Of these, we confirm 184
\!\!, listed in Table~\ref{tab:known_G22},
and show examples in Figure~\ref{fig:known_G22}.  In this table we revise the coordinates and sizes of the SNRs to enclose the SMGPS emission, but use the G22 source names.

Although some known SNRs in the sample do have MIR emission, the quality of this emission is very different from that of \hii\ regions.  For example, for SNR G11.2$-$0.3 shown in Figure~\ref{fig:known_G22}, there is bright 24\,\micron\ emission.  Compared to \hii\ regions, however, this source is lacking the $\sim\!10\micron$ emission that delineates a photodissociation region.  Therefore, one can visually distinguish between SNRs and \hii\ regions, even for those SNRs that are bright at 24\,\micron.

\begin{figure*}
        \centering
    \includegraphics[width=3.6in]{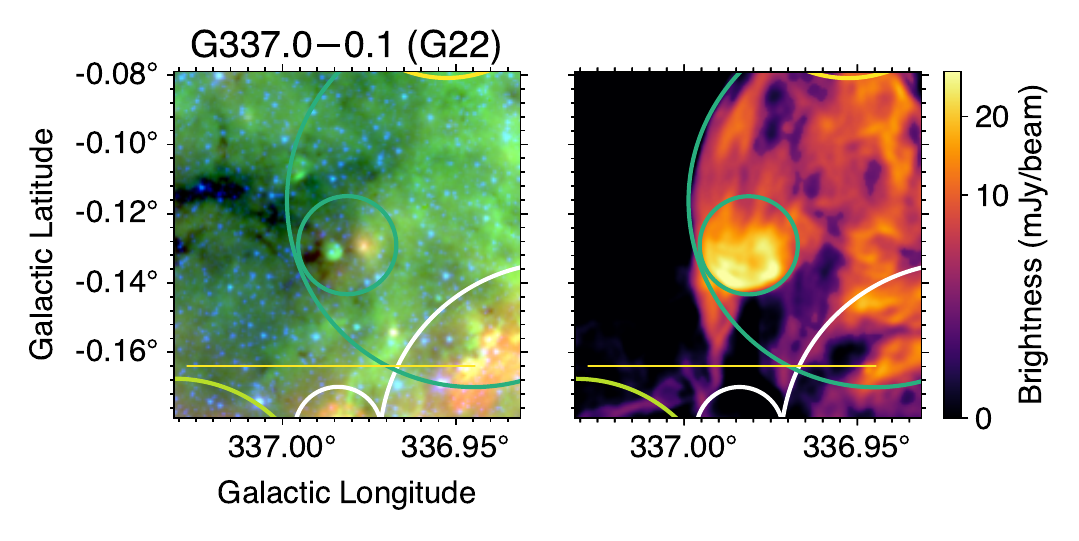}
    \includegraphics[width=3.6in]{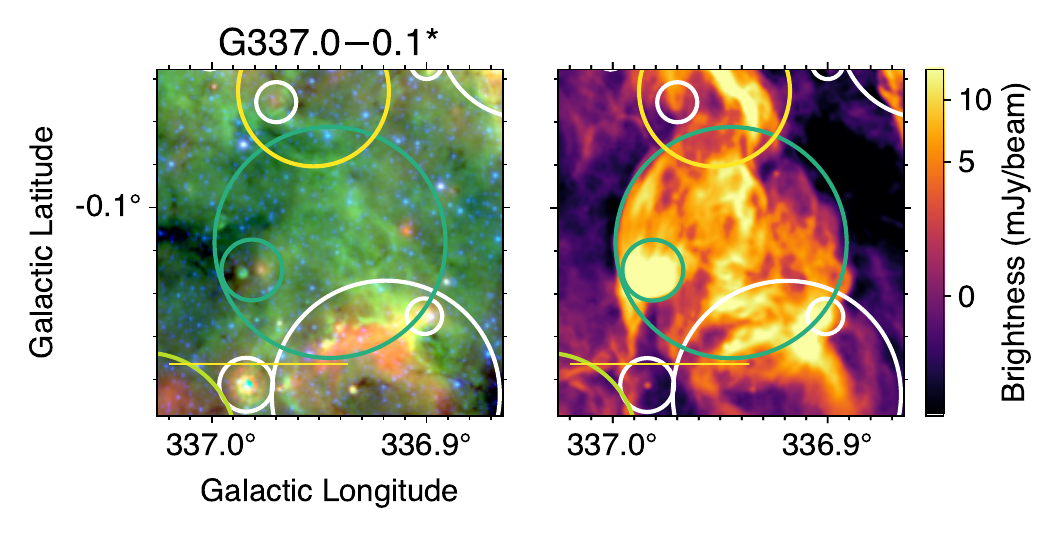}
    \caption{    
    Objects from the G22 catalog that we split into two entries, G337.0$-$0.1* (left panels) and G337.0$-$0.1 (right panels).  MIR data is in the left panels and SMGPS 1.3\,\ghz\ data are on the right.  The symbols have the same meaning as in Figure~\ref{fig:overview}, with the G22 SNRs being represented by blue-green circles at the centers of all panels.  Yellow scale bars at the lower left of each panel are $5\arcmin$ long.}
    \label{fig:G337}
\end{figure*}

We separate one G22 source into two SNRs.  The G22 source G337$-$0.1 refers to a small $\sim\!1\arcmin$ diameter object.  The larger shell, which was also called G337$-$0.1 in previous versions of the G22 catalog, is also detected in SMGPS data; we include the larger shell as a separate catalog entry using the name G337$-$0.1$^*$.  We show these two regions in Figure~\ref{fig:G337}.

\begin{figure*}
    \centering
    \includegraphics[width=3.6in]{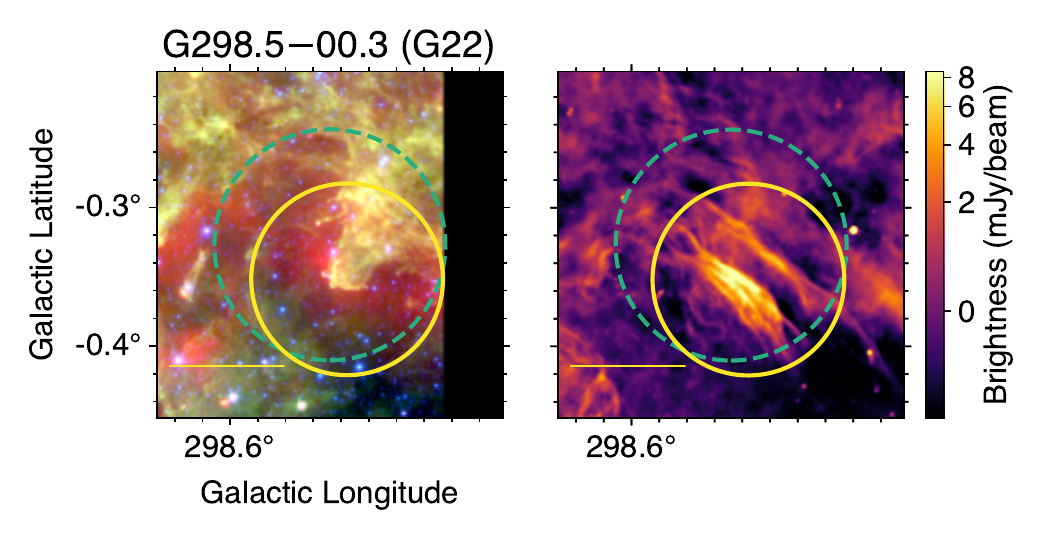}    \includegraphics[width=3.6in]{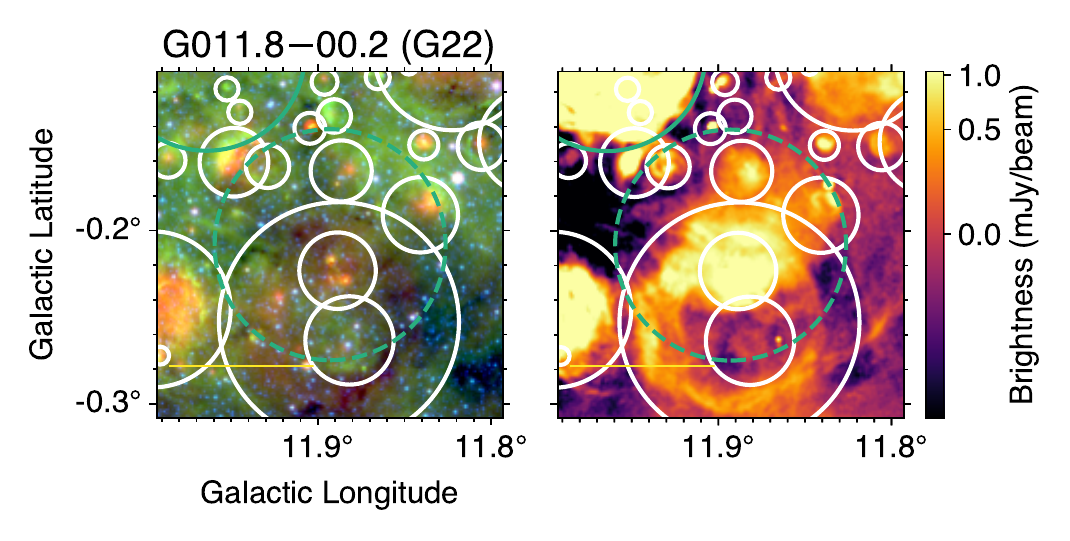}
    \caption{Objects from the G22 catalog that are unlikely to be true SNRs.  MIR data is in the left panels (Spitzer or WISE) and SMGPS 1.3\,\ghz\ data are on the right.  The symbols have the same meaning as in Figure~\ref{fig:overview}, with the G22 SNRs being indicated by blue-green dashed circles at the centers of all panels.  Yellow scale bars at the lower left of each panel are $5\arcmin$ long.}
    \label{fig:G22_bad}
\end{figure*}

We think that two G22 sources are unlikely to be true SNRs.  Although the source G298.5$-$0.3 is unambiguously detected in SMGPS data, it consists of two linear filaments and has a radio morphology inconsistent with the morphology of known SNRs.  It fits the criteria for ``unusual'' sources discussed later and is included there.  This source looks like a convincing SNR in low-resolution radio continuum emission from \citet{whiteoak96}, but the SMGPS data reveals its true morphology. The source G011.8$-$00.2 overlaps with the WISE Catalog \hii\ region G011.887$-$00.253.  We show these two sources in Figure~\ref{fig:G22_bad}.  

We detect SMGPS emission from 30
SNRCat objects that are not in the G22 catalog, out of 51 examined.  
We list the detected SNRCat objects in Table~\ref{tab:known_SNRCat}.  In this table we revise the coordinates and sizes of the sources to enclose the SMGPS emission, but use the catalog source names.

Of the detected sources, six are compact ($\sim\!30\arcsec$ or less) pulsar wind nebulae (PWNe; G267.0$-$1.0, G284.0$-$1.8, G336.4+0.2, G350.2$-$0.8, G18.0$-$0.7, and G36.0+0.1).  Two are long, linear PWNe (G319.9$-$0.7 and G337.5$-$0.1). Five are listed in SNRCat as being PWNe or candidates, but have extended radio continuum emission (G313.3+0.1, G313.6+0.3, G333.9+00.0, G348.9$-$0.4, and G29.4+0.1).  The remaining 17 sources have radio morphologies similar to the SNRs in G22.  Of these 17, 15 are listed in \citet{green14b} as being SNR candidates.

We suggest that nine of the examined SNRCat sources are not true SNRs.  Of these, four overlap with WISE Catalog \hii\ regions: G8.3$-$0.0 overlaps with
G008.306$-$00.084, G10.5$-$0.0 overlaps with G010.585$-$00.051, G11.0$-$1.0 overlaps with G011.182$-$01.063, and G14.3+0.1 overlaps with G014.303+00.140.  The radio emission from one, G354.4+0.0, consists of a broad radio filament and based on their MIR emission four appear to be thermal but are not confused with individual \hii\ regions: 
G284.2$-$0.4, G304.1$-$0.2, G21.9$-$0.1, and G026.6$-$0.1.  

Additionally, 12 SNRCat sources have no identifiable SMGPS emission: G285.1$-$00.5, G287.4+00.6, G323.9+00.0, G331.5$-$00.6, G332.5$-$00.3, G344.7+00.1, G350.2$-$00.8, G358.1+00.1, G358.3+00.2, G018.5$-$00.4, G032.73+0.15, G032.6+00.5, G044.5$-$00.2.  All except for G331.5$-$00.6 are listed as being PWNe or candidates, so their lack of SMGPS emission is unsurprising.  
\vskip 15pt
\topcaption{\label{tab:known_G22} Previously-Identified SNRs From G22}
\tablefirsthead{
\hline\hline
Name\tablefootmark{a} & GLong & GLat & Radius\\
 &  deg. &  deg. &  arcmin. \\
\hline}
\tablehead{%
\multicolumn{3}{@{}l}{Table \thetable, continued.}\\
\hline\hline
Name\tablefootmark{a} & GLong\tablefootmark{b} & GLat & Radius\\
 &  deg. &  deg. &  arcmin.\\
\hline}
\tabletail{\hline}
\tablelasttail{\hline}
\begin{xtabular}{lrrr}
G266.2$-$1.2  &  266.183  &  $-$1.193  &  51.5  \\ 
G279.0+1.1  &  278.958  &  0.761  &  67.2  \\ 
G284.3$-$1.8  &  284.279  &  $-$1.826  &  17.3  \\ 
G286.5$-$1.2  &  286.217  &  $-$1.104  &  23.2  \\ 
G289.7$-$0.3  &  289.707  &  $-$0.295  &  9.5  \\ 
G290.1$-$0.8  &  290.102  &  $-$0.754  &  9.6  \\ 
G291.0$-$0.1  &  291.033  &  $-$0.099  &  7.5  \\ 
G292.2$-$0.5  &  292.142  &  $-$0.532  &  9.9  \\ 
G293.8+0.6  &  293.766  &  0.608  &  11.8  \\ 
G294.1$-$0.0  &  294.125  &  $-$0.046  &  21.7  \\ 
G296.1$-$0.5  &  295.929  &  $-$0.547  &  33.1  \\ 
G296.7$-$0.9  &  296.661  &  $-$0.922  &  7.6  \\ 
G296.8$-$0.3  &  296.885  &  $-$0.350  &  10.3  \\ 
G298.6$-$0.0  &  298.573  &  $-$0.063  &  6.2  \\ 
G299.6$-$0.5  &  299.593  &  $-$0.476  &  8.0  \\ 
G301.4$-$1.0  &  301.432  &  $-$1.006  &  14.4  \\ 
G302.3+0.7  &  302.282  &  0.717  &  9.7  \\ 
G304.6+0.1  &  304.585  &  0.126  &  4.6  \\ 
G306.3$-$0.9  &  306.309  &  $-$0.890  &  2.2  \\ 
G308.1$-$0.7  &  308.122  &  $-$0.672  &  9.2  \\ 
G308.4$-$1.4  &  308.477  &  $-$1.373  &  8.2  \\ 
G308.8$-$0.1  &  308.860  &  $-$0.199  &  8.8  \\ 
G309.2$-$0.6  &  309.172  &  $-$0.698  &  8.3  \\ 
G309.8+0.0  &  309.779  &  0.006  &  13.6  \\ 
G310.6$-$0.3  &  310.618  &  $-$0.283  &  4.7  \\ 
G310.8$-$0.4  &  310.816  &  $-$0.475  &  5.8  \\ 
G311.5$-$0.3  &  311.524  &  $-$0.340  &  2.4  \\ 
G312.4$-$0.4  &  312.452  &  $-$0.423  &  17.6  \\ 
G315.4$-$0.3  &  315.469  &  $-$0.253  &  12.4  \\ 
G315.9$-$0.0  &  315.895  &  $-$0.015  &  10.0  \\ 
G316.3$-$0.0  &  316.272  &  $-$0.021  &  12.9  \\ 
G317.3$-$0.2  &  317.296  &  $-$0.258  &  6.8  \\ 
G318.2+0.1  &  318.148  &  0.161  &  28.3  \\ 
G318.9+0.4  &  318.897  &  0.411  &  16.0  \\ 
G320.4$-$1.2  &  320.382  &  $-$1.232  &  18.0  \\ 
G321.9$-$0.3  &  321.913  &  $-$0.312  &  15.9  \\ 
G321.9$-$1.1  &  321.929  &  $-$1.069  &  16.0  \\ 
G322.1+0.0  &  322.105  &  0.023  &  3.7  \\ 
G322.5$-$0.1  &  322.471  &  $-$0.101  &  8.0  \\ 
G323.5+0.1  &  323.481  &  0.072  &  6.9  \\ 
G323.7$-$1.0  &  323.591  &  $-$1.134  &  15.3  \\ 
G326.3$-$1.8  &  326.310  &  $-$1.714  &  18.5  \\ 
G327.1$-$1.1  &  327.139  &  $-$1.075  &  10.5  \\ 
G327.2$-$0.1  &  327.239  &  $-$0.129  &  2.9  \\ 
G327.4+0.4  &  327.276  &  0.520  &  11.4  \\ 
G327.4+1.0  &  327.385  &  1.002  &  7.4  \\ 
G328.4+0.2  &  328.412  &  0.225  &  3.3  \\ 
G329.7+0.4  &  329.722  &  0.374  &  20.2  \\ 
G330.2+1.0  &  330.155  &  0.977  &  5.5  \\ 
G332.0+0.2  &  331.989  &  0.173  &  5.9  \\ 
G332.4+0.1  &  332.387  &  0.127  &  7.3  \\ 
G332.4$-$0.4  &  332.423  &  $-$0.373  &  5.5  \\ 
G335.2+0.1  &  335.182  &  0.109  &  11.0  \\ 
G336.7+0.5  &  336.740  &  0.540  &  6.8  \\ 
G337.0$-$0.1  &  336.981  &  $-$0.129  &  0.8  \\ 
G337.2+0.1  &  337.184  &  0.065  &  1.5  \\ 
G337.2$-$0.7  &  337.210  &  $-$0.724  &  3.1  \\ 
G337.3+1.0  &  337.336  &  0.959  &  6.6  \\ 
G337.8$-$0.1  &  337.807  &  $-$0.094  &  3.6  \\ 
G338.1+0.4  &  338.076  &  0.424  &  7.6  \\ 
G338.3$-$0.0  &  338.327  &  $-$0.044  &  4.2  \\ 
G338.5+0.1  &  338.535  &  0.101  &  3.6  \\ 
G340.4+0.4  &  340.405  &  0.422  &  4.9  \\ 
G340.6+0.3  &  340.586  &  0.346  &  2.9  \\ 
G341.2+0.9  &  341.116  &  0.830  &  14.9  \\ 
G341.9$-$0.3  &  341.846  &  $-$0.306  &  4.2  \\ 
G342.0$-$0.2  &  341.956  &  $-$0.200  &  5.5  \\ 
G342.1+0.9  &  342.107  &  0.888  &  5.5  \\ 
G343.1$-$0.7  &  343.072  &  $-$0.668  &  12.0  \\ 
G344.7$-$0.1  &  344.682  &  $-$0.184  &  5.1  \\ 
G345.1$-$0.2  &  345.063  &  $-$0.204  &  3.5  \\ 
G345.1+0.2  &  345.155  &  0.244  &  6.6  \\ 
G345.7$-$0.2  &  345.734  &  $-$0.186  &  3.2  \\ 
G346.6$-$0.2  &  346.611  &  $-$0.218  &  4.9  \\ 
G347.3$-$0.5  &  347.362  &  $-$0.582  &  28.8  \\ 
G348.5+0.1  &  348.454  &  0.094  &  4.7  \\ 
G348.5$-$0.0  &  348.547  &  $-$0.003  &  5.1  \\ 
G348.7+0.3  &  348.646  &  0.399  &  5.4  \\ 
G348.8+1.1  &  348.860  &  1.126  &  6.5  \\ 
G349.2$-$0.1  &  349.149  &  $-$0.092  &  4.0  \\ 
G349.7+0.2  &  349.728  &  0.172  &  1.4  \\ 
G350.1$-$0.3  &  350.065  &  $-$0.325  &  2.7  \\ 
G351.2+0.1  &  351.263  &  0.164  &  4.8  \\ 
G351.7+0.8  &  351.710  &  0.819  &  8.1  \\ 
G351.9$-$0.9  &  351.926  &  $-$0.959  &  6.9  \\ 
G352.7$-$0.1  &  352.744  &  $-$0.125  &  4.1  \\ 
G353.3$-$1.1  &  353.514  &  $-$1.213  &  21.5  \\ 
G353.6$-$0.7  &  353.541  &  $-$0.670  &  16.7  \\ 
G354.1+0.1  &  354.166  &  0.053  &  4.5  \\ 
G354.8$-$0.8  &  354.864  &  $-$0.764  &  10.3  \\ 
G355.4+0.7  &  355.345  &  0.627  &  24.0  \\ 
G355.6$-$0.0  &  355.675  &  $-$0.078  &  3.9  \\ 
G356.3$-$0.3  &  356.312  &  $-$0.357  &  5.2  \\ 
G356.3$-$1.5  &  356.322  &  $-$1.535  &  10.1  \\ 
G357.7+0.3  &  357.706  &  0.324  &  13.8  \\ 
G357.7$-$0.1  &  357.714  &  $-$0.112  &  9.7  \\ 
G358.1+1.0  &  358.119  &  1.052  &  12.1  \\ 
G1.4$-$0.1  &  1.448  &  $-$0.148  &  5.6  \\ 
G1.9+0.3  &  1.871  &  0.323  &  0.9  \\ 
G3.1$-$0.6  &  3.061  &  $-$0.691  &  28.0  \\ 
G3.7$-$0.2  &  3.779  &  $-$0.288  &  7.0  \\ 
G3.8+0.3  &  3.860  &  0.393  &  10.9  \\ 
G5.4$-$1.2  &  5.378  &  $-$1.222  &  18.6  \\ 
G5.5+0.3  &  5.667  &  0.126  &  20.7  \\ 
G6.1+0.5  &  6.055  &  0.499  &  6.1  \\ 
G6.1+1.2  &  6.183  &  1.107  &  8.0  \\ 
G6.4$-$0.1  &  6.485  &  $-$0.102  &  19.0  \\ 
G6.5$-$0.4  &  6.521  &  $-$0.483  &  10.6  \\ 
G7.0$-$0.1  &  7.075  &  $-$0.102  &  9.1  \\ 
G7.2+0.2  &  7.204  &  0.177  &  4.9  \\ 
G8.9+0.4  &  8.778  &  0.477  &  18.9  \\ 
G8.7$-$0.1  &  8.828  &  $-$0.153  &  18.9  \\ 
G9.7$-$0.0  &  9.682  &  $-$0.077  &  5.0  \\ 
G9.8+0.6  &  9.781  &  0.571  &  7.5  \\ 
G9.9$-$0.8  &  9.974  &  $-$0.822  &  6.2  \\ 
G11.0$-$0.0  &  11.037  &  $-$0.053  &  4.5  \\ 
G11.1$-$0.7  &  11.130  &  $-$0.775  &  8.4  \\ 
G11.2$-$0.3  &  11.180  &  $-$0.348  &  2.8  \\ 
G11.1+0.1  &  11.186  &  0.128  &  5.1  \\ 
G11.4$-$0.1  &  11.383  &  $-$0.073  &  4.8  \\ 
G12.0$-$0.1  &  11.964  &  $-$0.097  &  3.4  \\ 
G12.2+0.3  &  12.262  &  0.305  &  3.4  \\ 
G12.5+0.2  &  12.580  &  0.217  &  2.7  \\ 
G12.7$-$0.0  &  12.712  &  $-$0.008  &  3.2  \\ 
G12.8$-$0.0  &  12.819  &  $-$0.023  &  1.5  \\ 
G13.1$-$0.5  &  13.078  &  $-$0.584  &  12.8  \\ 
G13.5+0.2  &  13.454  &  0.139  &  2.6  \\ 
G22.1$-$0.1  &  14.209  &  $-$0.076  &  11.3  \\ 
G15.4+0.1  &  15.427  &  0.161  &  7.3  \\ 
G15.5$-$0.1  &  15.506  &  $-$0.148  &  5.4  \\ 
G15.9+0.2  &  15.888  &  0.198  &  3.2  \\ 
G16.0$-$0.5  &  16.053  &  $-$0.479  &  11.7  \\ 
G16.7+0.1  &  16.735  &  0.082  &  2.6  \\ 
G17.0$-$0.0  &  17.016  &  $-$0.035  &  2.5  \\ 
G17.4$-$0.1  &  17.476  &  $-$0.117  &  5.0  \\ 
G18.1$-$0.1  &  18.150  &  $-$0.170  &  4.6  \\ 
G18.6$-$0.2  &  18.620  &  $-$0.280  &  3.4  \\ 
G18.8+0.3  &  18.770  &  0.390  &  10.3  \\ 
G18.9$-$1.1  &  18.950  &  $-$1.077  &  19.0  \\ 
G19.1+0.2  &  19.240  &  0.260  &  20.2  \\ 
G20.0$-$0.2  &  19.990  &  $-$0.190  &  7.3  \\ 
G21.0$-$0.4  &  21.030  &  $-$0.470  &  5.6  \\ 
G21.5$-$0.9  &  21.500  &  $-$0.885  &  1.5  \\ 
G21.6$-$0.8  &  21.640  &  $-$0.830  &  8.1  \\ 
G21.8$-$0.6  &  21.830  &  $-$0.520  &  15.5  \\ 
G22.7$-$0.2  &  22.710  &  $-$0.200  &  13.9  \\ 
G23.3$-$0.3  &  23.270  &  $-$0.330  &  16.8  \\ 
G24.7+0.6  &  24.550  &  0.630  &  15.2  \\ 
G24.7$-$0.6  &  24.860  &  $-$0.660  &  13.1  \\ 
G27.4+0.0  &  27.390  &  $-$0.010  &  3.1  \\ 
G27.8+0.6  &  27.700  &  0.630  &  23.4  \\ 
G28.3+0.2  &  28.365  &  0.209  &  6.5  \\ 
G28.6$-$0.1  &  28.610  &  $-$0.110  &  5.3  \\ 
G28.7$-$0.4  &  28.784  &  $-$0.437  &  6.0  \\ 
G29.6+0.1  &  29.564  &  0.110  &  3.1  \\ 
G29.7$-$0.3  &  29.710  &  $-$0.237  &  2.4  \\ 
G30.7+1.0  &  30.700  &  1.010  &  11.7  \\ 
G31.5$-$0.6  &  31.544  &  $-$0.660  &  10.7  \\ 
G31.9+0.0  &  31.870  &  0.017  &  4.1  \\ 
G32.1$-$0.9  &  32.130  &  $-$0.960  &  21.5  \\ 
G32.4+0.1  &  32.420  &  0.110  &  4.4  \\ 
G32.8$-$0.1  &  32.790  &  $-$0.040  &  11.5  \\ 
G33.2$-$0.6  &  33.180  &  $-$0.570  &  9.2  \\ 
G33.6+0.1  &  33.670  &  0.030  &  6.7  \\ 
G34.7$-$0.4  &  34.660  &  $-$0.400  &  19.2  \\ 
G35.6$-$0.4  &  35.590  &  $-$0.440  &  8.6  \\ 
G36.6$-$0.7  &  36.591  &  $-$0.818  &  6.5  \\ 
G38.7$-$1.3  &  38.756  &  $-$1.151  &  25.1  \\ 
G39.2$-$0.3  &  39.220  &  $-$0.320  &  4.5  \\ 
G40.5$-$0.5  &  40.520  &  $-$0.510  &  12.5  \\ 
G41.1$-$0.3  &  41.120  &  $-$0.314  &  2.7  \\ 
G41.5+0.4  &  41.450  &  0.410  &  8.5  \\ 
G42.0$-$0.1  &  41.960  &  $-$0.050  &  5.9  \\ 
G42.8+0.6  &  42.773  &  0.666  &  13.4  \\ 
G43.3$-$0.2  &  43.273  &  $-$0.182  &  3.1  \\ 
G45.7$-$0.4  &  45.610  &  $-$0.390  &  14.0  \\ 
G46.8$-$0.3  &  46.767  &  $-$0.280  &  10.0  \\ 
G49.2$-$0.7  &  49.185  &  $-$0.520  &  19.2  \\ 
G53.4+0.0  &  53.416  &  0.042  &  4.2  \\ 
G54.1+0.3  &  54.099  &  0.261  &  7.1  \\ 
G54.4$-$0.3  &  54.500  &  $-$0.270  &  25.0  \\ 
G55.0+0.3  &  54.935  &  $-$0.086  &  37.9  \\ 
G57.2+0.8  &  57.240  &  0.820  &  6.7  \\ 
G59.5+0.1  &  59.590  &  0.110  &  9.0  \\ 

\end{xtabular}
\tablefoot{\tablefoottext{a}{The source names listed in G22.} \tablefoottext{b}{Source centers and radii were determined by-eye from SMGPS data.}}

\vskip 15pt

\vskip 15pt
\topcaption{\label{tab:known_SNRCat} Previously-Identified Objects From SNRCat}
\tablefirsthead{
\hline\hline
Name\tablefootmark{a} & GLong & GLat & Radius\\
 &  deg. &  deg. &  arcmin. \\
\hline}
\tablehead{%
\multicolumn{3}{@{}l}{Table \thetable, continued.}\\
\hline\hline
Name\tablefootmark{a} & GLong\tablefootmark{b} & GLat & Radius\\
 &  deg. &  deg. &  arcmin.\\
\hline}
\tabletail{\hline}
\tablelasttail{\hline}
\begin{xtabular}{lrrr}
G267.0$-$1.0  &  266.967  &  $-$1.005  &  0.5  \\ 
G269.7+0.0  &  269.638  &  0.030  &  21.5  \\ 
G284.0$-$1.8  &  284.079  &  $-$1.880  &  0.4  \\ 
G291.0+0.1  &  291.011  &  0.079  &  13.3  \\ 
G296.6$-$0.4  &  296.608  &  $-$0.388  &  6.8  \\ 
G299.3$-$1.5  &  299.294  &  $-$1.521  &  18.5  \\ 
G310.9$-$0.3  &  310.887  &  $-$0.262  &  8.4  \\ 
G313.3+0.1  &  313.295  &  0.124  &  2.3  \\ 
G313.6+0.3  &  313.594  &  0.318  &  7.4  \\ 
G319.9$-$0.7  &  319.901  &  $-$0.674  &  5.1  \\ 
G322.7+0.1  &  322.684  &  0.078  &  7.3  \\ 
G322.9$-$0.0  &  322.886  &  $-$0.002  &  7.0  \\ 
G324.1+0.0  &  324.117  &  0.046  &  5.1  \\ 
G325.0$-$0.3  &  324.984  &  $-$0.315  &  2.1  \\ 
G330.7+0.1  &  330.718  &  0.116  &  7.4  \\ 
G333.9+0.0  &  333.863  &  $-$0.046  &  6.0  \\ 
G334.0$-$0.8  &  333.895  &  $-$0.850  &  10.1  \\ 
G336.4+0.2  &  336.386  &  0.191  &  0.5  \\ 
G336.7$-$0.3  &  336.658  &  $-$0.300  &  2.7  \\ 
G336.9$-$0.5  &  336.846  &  $-$0.550  &  8.5  \\ 
G337.5$-$0.1  &  337.528  &  $-$0.111  &  3.9  \\ 
G346.2$-$1.0  &  346.190  &  $-$0.984  &  4.0  \\ 
G348.9$-$0.4  &  348.960  &  $-$0.455  &  20.5  \\ 
G350.2$-$0.8  &  350.219  &  $-$0.842  &  0.6  \\ 
G354.1+0.3  &  354.055  &  0.281  &  6.4  \\ 
G5.7$-$0.1  &  5.699  &  $-$0.088  &  7.1  \\ 
G18.0$-$0.7  &  18.001  &  $-$0.691  &  0.3  \\ 
G20.4+0.1  &  20.460  &  0.150  &  5.4  \\ 
G29.4+0.1  &  29.369  &  0.101  &  5.3  \\ 
G36.0+0.1  &  36.008  &  0.057  &  0.2  \\ 

\end{xtabular}
\tablefoot{\tablefoottext{a}{The source names listed in SNRCat.}
\tablefoottext{b}{Source centers and radii were determined by-eye from SMGPS data.}}
\vskip 15pt


\subsection{Previously-Identified SNR Candidates\label{sec:known_cand}}
We examine the SMGPS images for radio continuum emission from the 170
previously-known SNR candidates in the SMGPS zone. We find unambiguous SMGPS emission consistent with that of known SNRs from 130
SNR candidates.  We list these detected SNR candidates in Table~\ref{tab:cand}.  In this table, as with the known SNRs, we revise the coordinates and sizes of the SNR candidates to enclose the SMGPS emission, but use the published source names.  These names are based on the published coordinates and have different numbers of significant digits between different authors.

We suggest that 13 previously-identified SNR candidates are misidentified.  Although detected in SMGPS data, a different classification for these sources is warranted.  Two are part of known SNRs (G6.4500$-$0.5583 is part of SNR G6.5$-$0.4 and G8.8583$-$0.2583 is part of G8.7$-$0.1). Three are confused with other SNR candidates (G005.989+0.019 with G006.199+0.157, G28.92+0.26 with G028.929+0.254, and G039.203+0.811 with G039.038+0.748). Five are confused with \hii\ regions (G331.8-0.0 with G331.834-00.00, G002.228+0.058 with G2.227+0.058, G022.951$-$0.311 with G22.951$-$0.311, G028.524+0.268 with G028.647+00.198, and G037.506+0.777 with G037.506+0.777).  Sources G324.4$-$0.2 and G327.1+0.9 \citep{ball23} are MIPSGAL bubbles, which are likely caused by an evolved star.  Finally, G339.6$-$0.6 appears to be a radio galaxy.  
 
Radiation will create ionized gas when impingent on a molecular cloud. Like in an \hii\ region, this ionized gas will emit radio continuum emission and the associated heated dust grains will emit in the MIR.  A low-density pathway through the interstellar medium would allow photons from a high-mass star formation region to irradiate clouds far afield, making this scenario possible whenever the ionizing flux in a region of the interstellar medium is high.  The observational signature of this scenario is MIR emission spatially coincident with radio continuum emission, similar to that of an \hii\ region, but the morphology will not have the generally circular symmetry of an \hii\ region. For six previously-identified SNR candidates, we suggest that the emission is actually thermal, caused by irradiation from nearby star formation regions (G26.13+0.13, G028.877+0.241, G28.88+0.41, G34.93-0.24, G39.56$-$0.32, and G043.070+0.558).  

Fifteen previously-identified SNR candidates appear to consist of nonthermal radio continuum emission (with no MIR), but the radio emission from each one does not form a cohesive structure (G005.106+0.332, G013.549+0.352, G013.626+0.299, G013.658$-$0.241, G016.126+0.690, G18.53$-$0.86, G19.13+0.90, G021.492$-$0.010, G022.177+0.314, G024.193+0.284, G27.39+0.24, G28.21+0.02, G030.362+0.623, G34.93$-$0.24, and G037.337+0.422).  Most of these consist of a single broad radio filament.  We suggest that these sources are poor SNR candidates that deserve future study.

While some previously-known SNR candidates are detected in the SMGPS, we decompose their emission into a number of new SNR candidates or find a dramatically different centroid.  We decompose the SNR candidate G328.6+0.0 \citep{ball23} into two regions.  We decompose SNR candidate G323.6$-$1.1 \citep{ball23} into two new SNR candidates and the known SNR G323.7$-$1.0.  Finally, we revise the position of G325.8+0.3 substantially.  These SNR candidates are not included in Table~\ref{tab:cand}.

Ten SNR candidates are not detected in SMGPS data: G350.7+0.6, G353.0+0.8,  G358.3$-$0.7, G18.9$-$1.2, G21.8+0.2, G24.0$-$0.3, G030.508+0.574, G32.73+0.15, G035.129$-$0.343, and G35.3$-$0.0.  We can make no judgement on the veracity of these classifications. Most (6 of the 10) sources have angular sizes $>30\arcmin$, which makes their detection difficult in SMGPS data. Two (G32.73+0.15 and G35.3$-$0.0) are in regions of the SMGPS with high noise.

\vskip 15pt
\topcaption{Previously-Identified SNR Candidates\label{tab:cand}}
\tablefirsthead{
\hline\hline
Name\tablefootmark{a} & GLong\tablefootmark{b} & GLat & Radius & Ref. \\
 &  deg. &  deg. &  arcmin. & \\
\hline}
\tablehead{%
\multicolumn{3}{@{}l}{Table \thetable, continued.}\\
\hline\hline
Name\tablefootmark{a} & GLong & GLat & Radius
& Ref. \\
 &  deg. &  deg. &  arcmin. & \\
\hline}
\tabletail{\hline}
\tablelasttail{\hline}
\begin{xtabular}{lrrrr}
G304.437$-$0.229  &  304.437  &  $-$0.229  &  7.2  &  14  \\ 
G317.5+0.9  &  317.550  &  0.953  &  16.3  &  13  \\ 
G320.6$-$0.9  &  320.636  &  $-$0.926  &  1.7  &  13  \\ 
G323.2$-$1.0  &  323.223  &  $-$0.977  &  4.2  &  13, 14  \\ 
G323.7+0.0  &  323.720  &  $-$0.019  &  2.0  &  14  \\ 
G324.1$-$0.2  &  324.099  &  $-$0.183  &  5.6  &  14  \\ 
G324.3+0.2  &  324.265  &  0.244  &  2.3  &  14  \\ 
G324.4$-$0.4  &  324.375  &  $-$0.369  &  8.0  &  14  \\ 
G324.7+0.0  &  324.660  &  $-$0.013  &  2.2  &  14  \\ 
G324.8$-$0.1  &  324.790  &  $-$0.100  &  11.9  &  14  \\ 
G325.0$-$0.5  &  324.945  &  $-$0.477  &  16.3  &  14  \\ 
G325.0+0.2  &  324.959  &  0.174  &  2.5  &  14  \\ 
G328.0+0.7  &  328.035  &  0.677  &  4.6  &  14  \\ 
SCO J170029$-$421309  &  343.874  &  0.032  &  2.2  &  17  \\ 
SCO J165948$-$420527  &  343.918  &  0.182  &  3.4  &  17  \\ 
G346.5$-$0.1  &  346.514  &  $-$0.080  &  6.3  &  16  \\ 
G349.1$-$0.3  &  349.104  &  $-$0.825  &  6.0  &  5  \\ 
G351.0$-$0.6  &  351.033  &  $-$0.627  &  6.0  &  5  \\ 
G351.4+0.4  &  351.409  &  0.484  &  4.4  &  5  \\ 
G351.4+0.2  &  351.470  &  0.205  &  7.6  &  5  \\ 
G351.9+0.1  &  351.918  &  0.148  &  8.3  &  5  \\ 
G001.949$-$0.100  &  1.949  &  $-$0.100  &  8.3  &  12  \\ 
G001.975$-$0.460  &  1.975  &  $-$0.460  &  4.9  &  12  \\ 
G002.228+0.058  &  2.228  &  0.058  &  1.1  &  12  \\ 
G002.276+0.399  &  2.276  &  0.399  &  6.3  &  12  \\ 
G002.910$-$0.183  &  2.910  &  $-$0.183  &  12.9  &  12  \\ 
G003.101$-$0.093  &  3.101  &  $-$0.093  &  2.5  &  12  \\ 
G003.103+0.110  &  3.103  &  0.110  &  1.4  &  12  \\ 
G4.20$-$0.30  &  4.176  &  $-$0.261  &  4.0  &  1  \\ 
G004.493$-$0.391  &  4.493  &  $-$0.391  &  6.9  &  12  \\ 
G004.571$-$0.244  &  4.571  &  $-$0.244  &  4.2  &  12  \\ 
G005.161$-$0.321  &  5.161  &  $-$0.321  &  5.6  &  12  \\ 
G005.364$-$0.705  &  5.364  &  $-$0.705  &  10.0  &  12  \\ 
G005.378$-$0.280  &  5.378  &  $-$0.280  &  2.2  &  12  \\ 
G005.378+0.347  &  5.378  &  0.347  &  6.2  &  12  \\ 
G005.762+0.515  &  5.762  &  0.515  &  5.8  &  12  \\ 
G006.118+0.387  &  6.118  &  0.387  &  9.7  &  12  \\ 
G6.31+0.54  &  6.247  &  0.458  &  10.5  &  2  \\ 
G7.4+0.3  &  7.427  &  0.331  &  6.8  &  5  \\ 
G008.040+0.566  &  8.050  &  0.566  &  5.0  &  12  \\ 
G11.5500+0.3333  &  11.550  &  0.339  &  3.4  &  3  \\ 
G013.500+0.074  &  13.500  &  0.074  &  1.9  &  12  \\ 
G013.652+0.259  &  13.652  &  0.259  &  2.5  &  12  \\ 
G014.524+0.140  &  14.524  &  0.140  &  14.5  &  12  \\ 
G015.862+0.522  &  15.862  &  0.522  &  1.9  &  12  \\ 
G016.021+0.746  &  16.021  &  0.746  &  6.3  &  12  \\ 
G016.956$-$0.933  &  16.956  &  $-$0.933  &  7.4  &  12  \\ 
G017.434+0.273  &  17.434  &  0.273  &  2.1  &  12  \\ 
G017.593+0.237  &  17.593  &  0.237  &  1.5  &  12  \\ 
G017.620+0.086  &  17.620  &  0.086  &  3.0  &  12  \\ 
G17.80$-$0.02  &  17.810  &  $-$0.018  &  4.3  &  4  \\ 
G018.393$-$0.816  &  18.393  &  $-$0.816  &  2.6  &  12  \\ 
G18.76$-$0.07  &  18.761  &  $-$0.074  &  1.2  &  2, 4  \\ 
G019.481$-$0.108  &  19.481  &  $-$0.108  &  7.4  &  12  \\ 
G19.75$-$0.69  &  19.746  &  $-$0.694  &  12.2  &  4, 5  \\ 
G019.751+0.202  &  19.751  &  0.202  &  7.9  &  12  \\ 
G19.96$-$0.33  &  19.960  &  $-$0.330  &  5.9  &  4  \\ 
G020.195+0.028  &  20.195  &  0.028  &  3.6  &  12  \\ 
G20.26$-$0.86  &  20.286  &  $-$0.858  &  8.6  &  4  \\ 
G021.596$-$0.179  &  21.596  &  $-$0.179  &  10.4  &  12  \\ 
G021.684+0.129  &  21.684  &  0.129  &  3.8  &  12  \\ 
G021.861+0.169  &  21.861  &  0.169  &  2.7  &  12  \\ 
G022.045$-$0.028  &  22.045  &  $-$0.028  &  7.7  &  12  \\ 
G22.32+0.11  &  22.265  &  0.052  &  10.1  &  3, 4  \\ 
G23.11+0.19  &  23.116  &  0.186  &  11.7  &  4, 5, 6  \\ 
G23.85$-$0.18  &  23.855  &  $-$0.182  &  2.2  &  4  \\ 
G023.973+0.510  &  23.973  &  0.510  &  4.9  &  12  \\ 
G024.062$-$0.808  &  24.062  &  $-$0.808  &  2.0  &  12  \\ 
G25.49+0.01  &  25.482  &  0.009  &  6.5  &  4, 7  \\ 
G26.04$-$0.42  &  25.971  &  $-$0.392  &  11.4  &  4  \\ 
G26.53+0.07  &  26.530  &  0.070  &  11.2  &  4  \\ 
G26.75+0.73  &  26.753  &  0.693  &  7.2  &  4  \\ 
G27.06+0.04  &  27.025  &  0.067  &  9.0  &  3, 4, 8  \\ 
G27.18+0.30  &  27.179  &  0.303  &  1.3  &  4  \\ 
G27.24$-$0.14  &  27.240  &  $-$0.140  &  6.1  &  4  \\ 
G27.47+0.25  &  27.461  &  0.241  &  1.7  &  4  \\ 
G27.78$-$0.33  &  27.776  &  $-$0.325  &  3.3  &  4  \\ 
G28.22$-$0.09  &  28.216  &  $-$0.087  &  1.7  &  4  \\ 
G28.33+0.06  &  28.330  &  0.060  &  3.2  &  4  \\ 
G28.56+0.00  &  28.557  &  $-$0.012  &  2.3  &  3, 4, 8  \\ 
G28.64+0.20  &  28.640  &  0.200  &  11.4  &  3, 4  \\ 
G028.870+0.616  &  28.870  &  0.616  &  2.0  &  12  \\ 
G028.929+0.254  &  28.929  &  0.254  &  2.2  &  12  \\ 
G029.329+0.280  &  29.329  &  0.280  &  2.5  &  12  \\ 
G29.41$-$0.18  &  29.410  &  $-$0.180  &  7.5  &  4  \\ 
G030.303+0.128  &  30.303  &  0.128  &  1.0  &  12  \\ 
G030.375+0.424  &  30.375  &  0.424  &  2.3  &  12  \\ 
G31.22$-$0.02  &  31.256  &  $-$0.041  &  3.3  &  4  \\ 
G31.93+0.16  &  31.937  &  0.171  &  2.3  &  4  \\ 
G32.22$-$0.21  &  32.226  &  $-$0.210  &  3.2  &  4  \\ 
G032.458$-$0.112  &  32.458  &  $-$0.112  &  1.8  &  12  \\ 
G33.85+0.06  &  33.847  &  0.062  &  0.5  &  4  \\ 
G034.524$-$0.761  &  34.524  &  $-$0.761  &  2.7  &  12  \\ 
G034.619+0.240  &  34.619  &  0.240  &  3.0  &  12  \\ 
G36.66$-$0.50  &  36.660  &  $-$0.500  &  8.2  &  4  \\ 
G36.68$-$0.14  &  36.671  &  $-$0.137  &  9.5  &  4  \\ 
G036.839$-$0.433  &  36.839  &  $-$0.433  &  2.6  &  12  \\ 
G036.851$-$0.246  &  36.851  &  $-$0.246  &  1.5  &  12  \\ 
G36.90+0.49  &  36.902  &  0.488  &  3.8  &  4  \\ 
G37.62$-$0.22  &  37.621  &  $-$0.219  &  2.2  &  4  \\ 
G037.672$-$0.501  &  37.672  &  $-$0.501  &  2.8  &  12  \\ 
G37.88+0.32  &  37.897  &  0.289  &  13.8  &  4  \\ 
G38.17+0.09  &  38.138  &  0.056  &  10.5  &  4  \\ 
G38.62$-$0.24  &  38.620  &  $-$0.240  &  2.5  &  4  \\ 
G38.68$-$0.43  &  38.694  &  $-$0.453  &  2.6  &  4  \\ 
G38.83$-$0.01  &  38.838  &  $-$0.012  &  1.3  &  4  \\ 
G039.539+0.366  &  39.539  &  0.366  &  4.5  &  12  \\ 
G041.510$-$0.534  &  41.510  &  $-$0.534  &  16.0  &  12  \\ 
G041.625+0.261  &  41.625  &  0.241  &  4.7  &  12  \\ 
G41.95$-$0.18  &  41.950  &  $-$0.180  &  7.0  &  4  \\ 
G42.62+0.14  &  42.623  &  0.174  &  5.1  &  4  \\ 
G042.711$-$0.272  &  42.711  &  $-$0.272  &  9.1  &  12  \\ 
G043.023+0.726  &  43.011  &  0.746  &  6.5  &  12  \\ 
G043.502+0.667  &  43.502  &  0.667  &  10.1  &  15  \\ 
G044.076+0.127  &  44.076  &  0.127  &  8.0  &  12  \\ 
G46.18$-$0.02  &  46.155  &  $-$0.020  &  6.5  &  4  \\ 
G46.54$-$0.03  &  46.542  &  0.004  &  8.2  &  4  \\ 
G47.15+0.73  &  47.150  &  0.730  &  0.8  &  4  \\ 
G047.741$-$0.971  &  47.741  &  $-$0.971  &  9.8  &  12  \\ 
G048.875+0.174  &  48.875  &  0.174  &  6.6  &  12  \\ 
G051.061+0.563  &  51.061  &  0.563  &  4.7  &  12  \\ 
G51.21+0.11  &  51.209  &  0.113  &  14.9  &  4, 8, 9, 10  \\ 
G52.37$-$0.70  &  52.381  &  $-$0.725  &  18.9  &  4, 10  \\ 
G53.07+0.49  &  53.071  &  0.490  &  0.9  &  4, 10  \\ 
G53.84$-$0.75  &  53.840  &  $-$0.750  &  18.7  &  4, 10  \\ 
G56.56$-$0.75  &  56.502  &  $-$0.774  &  13.1  &  4  \\ 
G57.12+0.35  &  57.114  &  0.331  &  13.5  &  4  \\ 
G58.70$-$0.31  &  58.697  &  $-$0.314  &  4.0  &  4  \\ 
G59.46+0.83  &  59.455  &  0.819  &  3.8  &  4  \\ 
G059.834$-$0.405  &  59.834  &  $-$0.405  &  10.5  &  12  \\ 

\end{xtabular}
\tablefoot{\tablefoottext{a}{The source names in the original publications.}
\tablefoottext{b}{Source centers and radii were determined by-eye from SMGPS data.}}
\tablebib{
(1)~\citet{trushkin01}; (2)~\citet{brogan06}; (3)~\citet{helfand06}; (4)~\citet{anderson17}; (5)~\citet{hurley-walker19}; (6)~\citet{maxted19}; (7)~\citet{bamba03}; (8)~\citet{dokara18}; (9)~\citet{sidorin14}; (10)~\citet{driessen18}; (11)~\citet{supan18}; (12)~\citet{dokara21}; (13)~\citet{whiteoak96}; (14)~\citet{ball23}; (15)~\citet{sushch17}; (16)~\citet{kaplan02}; (17)~\citet{gaensler01};
}
\vskip 15pt

\subsection{Newly-Identified SNR Candidates\label{sec:cand}}
By visually inspecting SMGPS and MIR data, we identify 237
new SNR candidates.  All 237
SNR candidates have SMGPS 1.3\,\ghz\ continuum emission and a deficiency of MIR emission.

For each candidate, we assign a reliability criterion that indicates our confidence that the object is a true SNR ranging from ``I'' (high reliability) to ``III'' (low reliability)\footnote{See \citet{brogan06} and \citet{hurley-walker19}, who used the same nomenclature with slightly different definitions.}.  Sources with a reliability criterion of ``I'' have clear SMGPS emission that is not confused with that of other SNRs or \hii\ regions, with a morphology that is similar to that of known SNRs.  These are most likely to have a shell morphology.  Sources with a reliability criterion of ``II'' are either somewhat confused with that of other SNRs or \hii\ regions, or which have a morphology that is slightly ambiguous.  They may be cospatial with MIR emission, but we cannot be sure if this emission comes from the radio source or another region along the line of sight. Class II sources may also be incomplete shells.  Sources with a reliability criterion of ``III'' are either badly confused with other SNRs or \hii\ regions, have spatially coincident radio and MIR emission that may indicate that at least some of the radio emission is thermal, or are incomplete shells.  We assign 35
\!\!\% of the newly-discovered SNR candidates a reliability criterion of ``I,'' 44
\!\!\% of ``II'', and 22
\!\!\% as ``III.'' 

\begin{figure*}
    \centering
    \includegraphics[width=3.6in]{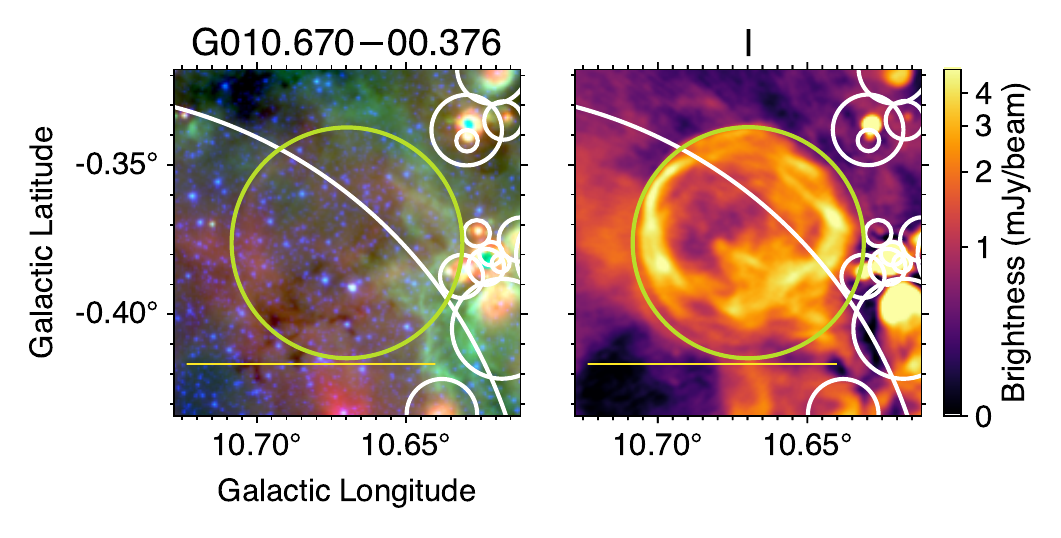}
    \includegraphics[width=3.6in]{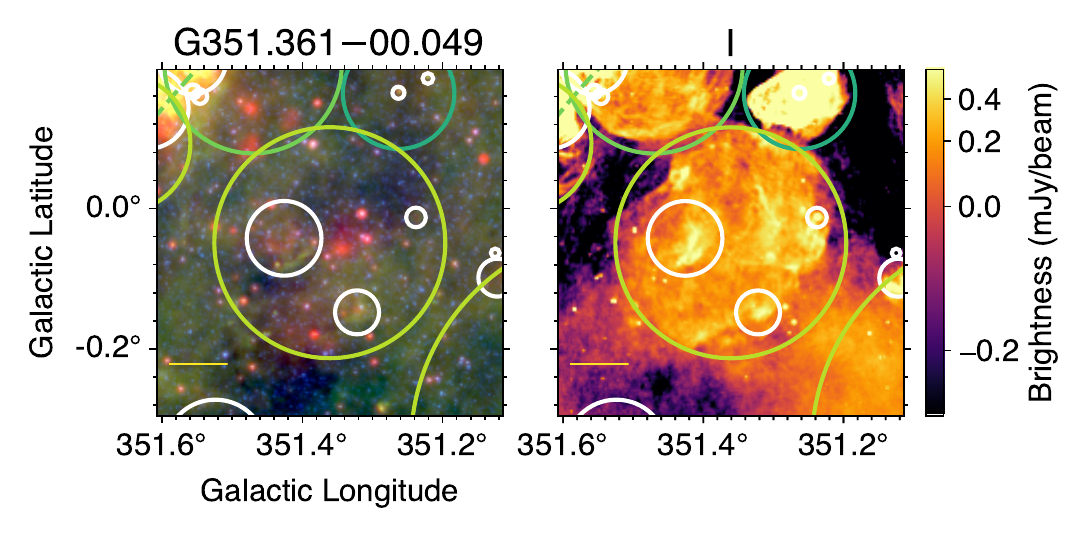}
    \includegraphics[width=3.6in]{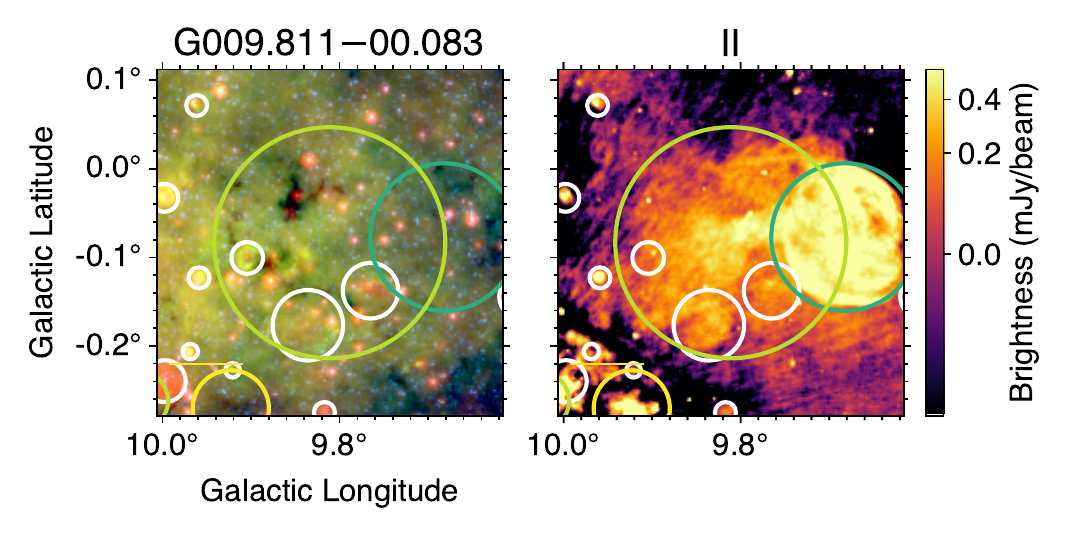}
    \includegraphics[width=3.6in]{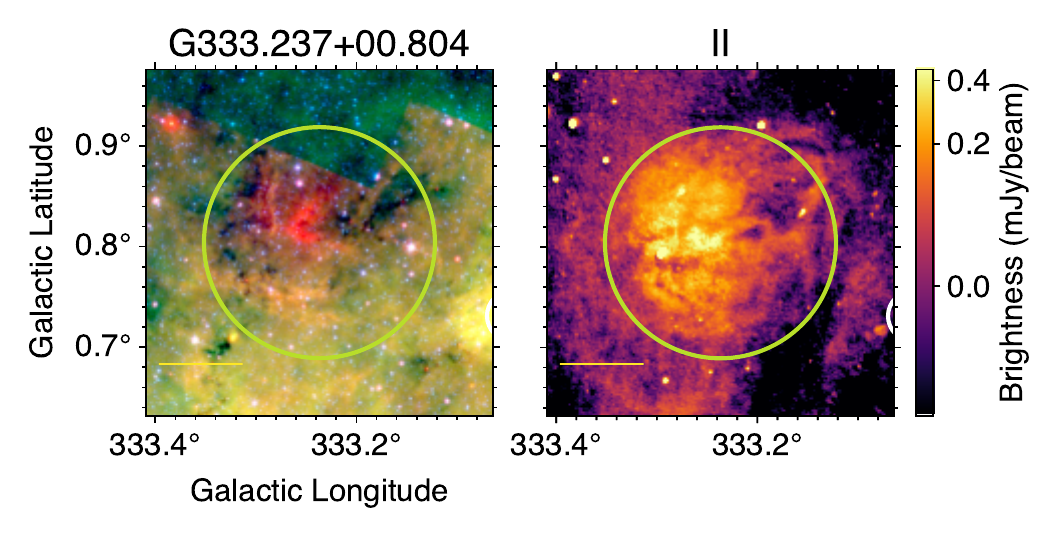}

    \includegraphics[width=3.6in]{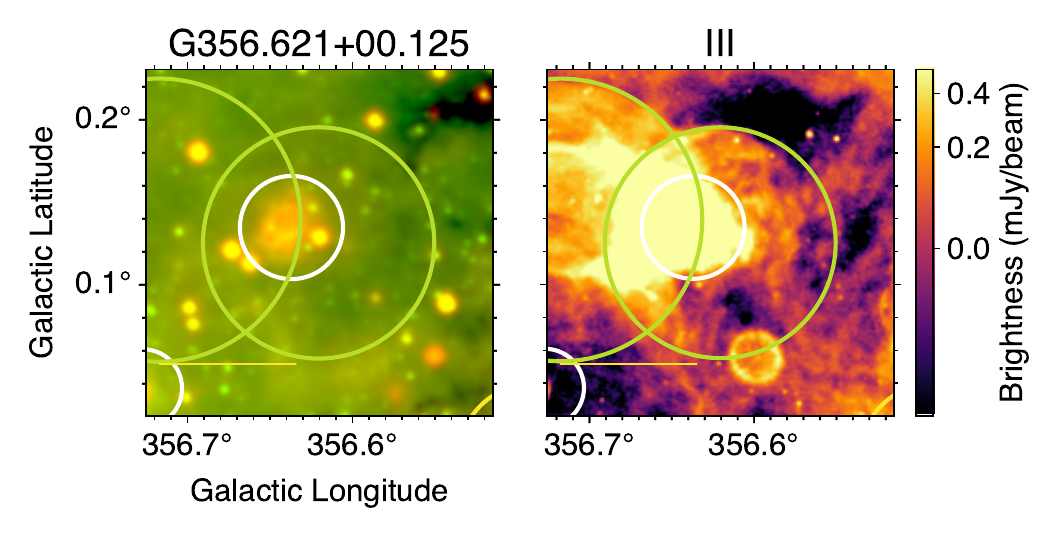}
    \includegraphics[width=3.6in]{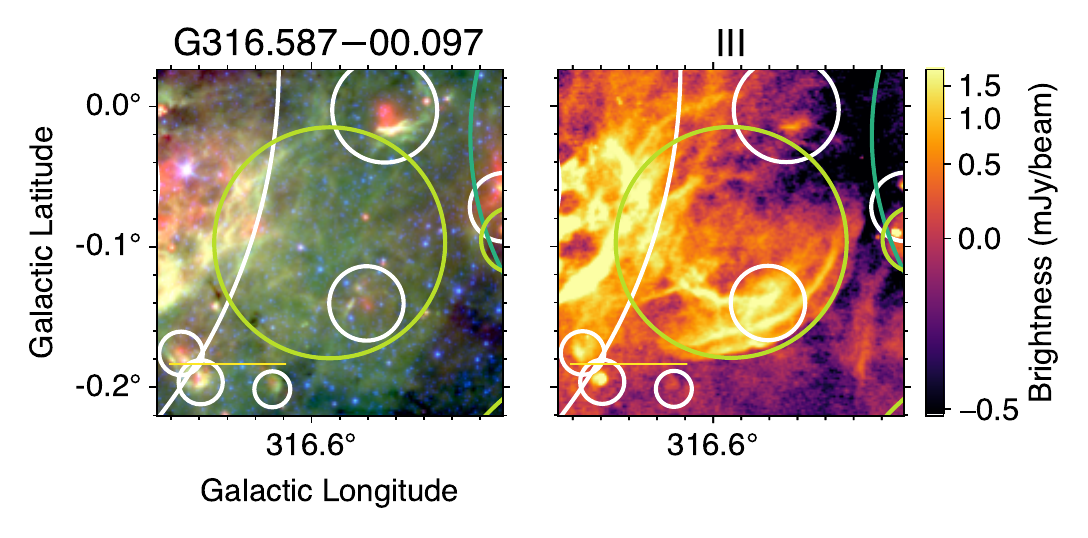}
    
    \caption{Example newly-discovered SNR candidates.  MIR data is in the left panels and SMGPS 1.3\,\ghz\ data are on the right.  The symbols have the same meaning as in Figure~\ref{fig:overview}, with the candidate SNRs being represented by light green circles at the centers of all panels.  On the top of the panels showing SMGPS images we list our reliability classification, with ``I'' being the most reliable and ``III'' the least.  Yellow scale bars at the lower left of each panel are $5\arcmin$ long.}
    \label{fig:new}
\end{figure*}

We give the parameters of the new SNR candidates in Table~\ref{tab:new} and show two examples of each reliability criterion in Figure~\ref{fig:new}.

\vskip 15pt
\begingroup\small
\topcaption{\label{tab:new}SMGPS SNR Candidates}
\tablefirsthead{
\hline\hline
Name & GLong & GLat & Radius & Reliability\tablefootmark{a} \\
 &  deg. &  deg. &  arcmin. & \\
\hline}
\tablehead{%
\multicolumn{3}{@{}l}{Table \thetable, continued.}\\
\hline\hline
Name & GLong & GLat & Radius & Reliability\tablefootmark{a} \\
 &  deg. &  deg. &  arcmin. & \\
\hline}
\tabletail{\hline}
\tablelasttail{\hline}
\begin{xtabular}{lrrrc}
G257.408$-$00.162  &  257.408  &  $-$0.162  &  35.6  &  I  \\ 
G259.683$-$01.235  &  259.683  &  $-$1.235  &  7.7  &  I  \\ 
G275.256$-$00.979  &  275.256  &  $-$0.979  &  17.2  &  I  \\ 
G275.986$-$01.077  &  275.986  &  $-$1.077  &  4.8  &  I  \\ 
G276.229+00.419  &  276.229  &  0.419  &  1.5  &  II  \\ 
G277.162+00.396  &  277.162  &  0.396  &  4.8  &  II  \\ 
G278.924$-$01.195  &  278.924  &  $-$1.195  &  24.1  &  II  \\ 
G280.151+00.206  &  280.151  &  0.206  &  12.8  &  I  \\ 
G281.188$-$00.053  &  281.188  &  $-$0.053  &  29.1  &  I  \\ 
G282.674$-$00.737  &  282.674  &  $-$0.737  &  10.4  &  I  \\ 
G283.234$-$00.945  &  283.234  &  $-$0.945  &  6.4  &  II  \\ 
G283.429$-$01.414  &  283.429  &  $-$1.414  &  10.4  &  II  \\ 
G283.629$-$00.933  &  283.629  &  $-$0.933  &  10.2  &  II  \\ 
G283.849$-$01.431  &  283.849  &  $-$1.431  &  9.9  &  III  \\ 
G286.667+00.909  &  286.667  &  0.909  &  6.5  &  II  \\ 
G286.825+00.025  &  286.825  &  0.025  &  21.2  &  II  \\ 
G288.824$-$00.155  &  288.824  &  $-$0.155  &  2.7  &  II  \\ 
G288.830$-$01.366  &  288.830  &  $-$1.366  &  20.0  &  III  \\ 
G288.863$-$00.046  &  288.863  &  $-$0.046  &  20.6  &  III  \\ 
G289.150$-$00.890  &  289.150  &  $-$0.890  &  15.6  &  I  \\ 
G290.003$-$01.510  &  290.003  &  $-$1.510  &  4.3  &  II  \\ 
G290.575+00.465  &  290.575  &  0.465  &  15.0  &  II  \\ 
G292.313+00.587  &  292.313  &  0.587  &  9.5  &  II  \\ 
G292.472+00.167  &  292.472  &  0.167  &  0.9  &  III  \\ 
G295.855$-$01.388  &  295.855  &  $-$1.388  &  0.9  &  III  \\ 
G297.038$-$01.010  &  297.038  &  $-$1.010  &  31.2  &  II  \\ 
G299.680$-$00.015  &  299.680  &  $-$0.015  &  12.0  &  I  \\ 
G302.324$-$01.300  &  302.324  &  $-$1.300  &  6.0  &  II  \\ 
G303.365$-$00.050  &  303.365  &  $-$0.050  &  21.3  &  III  \\ 
G304.071+01.227  &  304.071  &  1.227  &  3.5  &  II  \\ 
G304.226$-$00.504  &  304.226  &  $-$0.504  &  5.3  &  I  \\ 
G306.195$-$00.520  &  306.195  &  $-$0.520  &  1.9  &  II  \\ 
G306.243$-$00.876  &  306.243  &  $-$0.876  &  13.4  &  III  \\ 
G307.050$-$00.691  &  307.050  &  $-$0.691  &  31.4  &  I  \\ 
G307.510$-$00.954  &  307.510  &  $-$0.954  &  11.1  &  I  \\ 
G307.930+00.134  &  307.930  &  0.134  &  13.8  &  I  \\ 
G308.309$-$00.189  &  308.309  &  $-$0.189  &  5.9  &  II  \\ 
G308.730+01.380  &  308.730  &  1.380  &  10.8  &  I  \\ 
G308.830$-$00.544  &  308.830  &  $-$0.544  &  15.6  &  I  \\ 
G309.202$-$00.118  &  309.202  &  $-$0.118  &  6.7  &  I  \\ 
G310.275$-$00.171  &  310.275  &  $-$0.171  &  5.2  &  I  \\ 
G310.457$-$00.607  &  310.457  &  $-$0.607  &  12.4  &  I  \\ 
G310.702$-$00.573  &  310.702  &  $-$0.573  &  1.2  &  II  \\ 
G310.955$-$00.565  &  310.955  &  $-$0.565  &  2.6  &  II  \\ 
G312.273$-$00.673  &  312.273  &  $-$0.673  &  9.7  &  I  \\ 
G312.504$-$00.018  &  312.504  &  $-$0.018  &  1.1  &  II  \\ 
G312.969$-$00.095  &  312.969  &  $-$0.095  &  8.1  &  I  \\ 
G313.499$-$00.525  &  313.499  &  $-$0.525  &  7.9  &  I  \\ 
G314.338$-$00.204  &  314.338  &  $-$0.204  &  12.8  &  I  \\ 
G314.620+00.256  &  314.620  &  0.256  &  12.0  &  III  \\ 
G315.312$-$00.319  &  315.312  &  $-$0.319  &  7.1  &  I  \\ 
G315.715$-$00.227  &  315.715  &  $-$0.227  &  0.9  &  II  \\ 
G315.501$-$00.584  &  315.501  &  $-$0.584  &  1.6  &  II  \\ 
G315.620+00.332  &  315.620  &  0.332  &  2.6  &  II  \\ 
G315.905$-$00.817  &  315.905  &  $-$0.817  &  13.2  &  II  \\ 
G316.349$-$00.340  &  316.349  &  $-$0.340  &  10.5  &  I  \\ 
G316.456$-$00.095  &  316.456  &  $-$0.095  &  1.4  &  II  \\ 
G316.587$-$00.097  &  316.587  &  $-$0.097  &  4.9  &  III  \\ 
G316.704+00.399  &  316.704  &  0.399  &  7.3  &  I  \\ 
G318.033$-$00.935  &  318.033  &  $-$0.935  &  1.5  &  II  \\ 
G318.072$-$00.394  &  318.072  &  $-$0.394  &  4.0  &  II  \\ 
G318.602$-$00.884  &  318.602  &  $-$0.884  &  9.2  &  II  \\ 
G318.871$-$00.468  &  318.871  &  $-$0.468  &  9.0  &  I  \\ 
G319.426$-$00.691  &  319.426  &  $-$0.691  &  9.4  &  I  \\ 
G320.762$-$00.351  &  320.762  &  $-$0.351  &  5.0  &  I  \\ 
G320.944$-$01.289  &  320.944  &  $-$1.289  &  7.4  &  III  \\ 
G321.311$-$00.862  &  321.311  &  $-$0.862  &  3.5  &  I  \\ 
G321.455+00.408  &  321.455  &  0.408  &  6.4  &  III  \\ 
G322.462+00.084  &  322.462  &  0.084  &  9.7  &  III  \\ 
G322.644$-$00.694  &  322.644  &  $-$0.694  &  15.0  &  I  \\ 
G322.762$-$00.423  &  322.762  &  $-$0.423  &  9.5  &  I  \\ 
G323.551$-$00.190  &  323.551  &  $-$0.190  &  27.2  &  I  \\ 
G323.552$-$00.790  &  323.552  &  $-$0.790  &  15.6  &  I  \\ 
G323.884$-$01.128  &  323.884  &  $-$1.128  &  14.6  &  I  \\ 
G324.009$-$01.295  &  324.009  &  $-$1.295  &  1.2  &  III  \\ 
G324.505+00.128  &  324.505  &  0.128  &  3.5  &  III  \\ 
G324.923$-$00.571  &  324.923  &  $-$0.571  &  5.7  &  II  \\ 
G325.502+00.293  &  325.502  &  0.293  &  29.1  &  II  \\ 
G326.622$-$00.558  &  326.622  &  $-$0.558  &  28.8  &  III  \\ 
G328.225$-$00.664  &  328.225  &  $-$0.664  &  5.5  &  I  \\ 
G328.346$-$00.480  &  328.346  &  $-$0.480  &  2.8  &  II  \\ 
G328.479$-$00.108  &  328.479  &  $-$0.108  &  9.3  &  I  \\ 
G328.703+00.016  &  328.703  &  0.016  &  9.3  &  I  \\ 
G330.628$-$00.284  &  330.628  &  $-$0.284  &  7.7  &  I  \\ 
G331.635+00.247  &  331.635  &  0.247  &  14.7  &  I  \\ 
G332.478+00.083  &  332.478  &  0.083  &  3.9  &  I  \\ 
G333.237+00.804  &  333.237  &  0.804  &  6.9  &  II  \\ 
G333.480+00.010  &  333.480  &  0.010  &  7.1  &  II  \\ 
G334.038+00.040  &  334.038  &  0.040  &  5.9  &  I  \\ 
G334.834$-$00.044  &  334.834  &  $-$0.044  &  0.9  &  II  \\ 
G334.929$-$00.464  &  334.929  &  $-$0.464  &  4.2  &  I  \\ 
G334.938$-$00.319  &  334.938  &  $-$0.319  &  1.8  &  II  \\ 
G334.991$-$00.290  &  334.991  &  $-$0.290  &  1.0  &  III  \\ 
G335.179+00.525  &  335.179  &  0.525  &  7.0  &  II  \\ 
G335.341$-$00.189  &  335.341  &  $-$0.189  &  1.7  &  II  \\ 
G335.386+00.252  &  335.386  &  0.252  &  5.9  &  II  \\ 
G336.321$-$00.748  &  336.321  &  $-$0.748  &  1.0  &  II  \\ 
G337.031$-$00.210  &  337.031  &  $-$0.210  &  2.5  &  III  \\ 
G337.069$-$00.340  &  337.069  &  $-$0.340  &  1.3  &  III  \\ 
G337.109$-$00.066  &  337.109  &  $-$0.066  &  1.3  &  II  \\ 
G337.167+00.332  &  337.167  &  0.332  &  2.5  &  II  \\ 
G338.763+00.374  &  338.763  &  0.374  &  11.5  &  III  \\ 
G339.169$-$00.023  &  339.169  &  $-$0.023  &  1.2  &  III  \\ 
G339.190$-$00.266  &  339.190  &  $-$0.266  &  6.5  &  I  \\ 
G339.304+00.038  &  339.304  &  0.038  &  5.6  &  III  \\ 
G339.446$-$00.151  &  339.446  &  $-$0.151  &  1.1  &  I  \\ 
G339.686$-$00.182  &  339.686  &  $-$0.182  &  4.2  &  I  \\ 
G340.132$-$00.069  &  340.132  &  $-$0.069  &  3.1  &  II  \\ 
G340.763$-$00.463  &  340.763  &  $-$0.463  &  7.4  &  II  \\ 
G341.137$-$01.237  &  341.137  &  $-$1.237  &  2.1  &  II  \\ 
G341.427$-$00.393  &  341.427  &  $-$0.393  &  4.2  &  II  \\ 
G341.456$-$00.176  &  341.456  &  $-$0.176  &  7.8  &  I  \\ 
G341.515+00.174  &  341.515  &  0.174  &  13.7  &  II  \\ 
G341.756$-$00.324  &  341.756  &  $-$0.324  &  5.8  &  I  \\ 
G341.830$-$00.591  &  341.830  &  $-$0.591  &  14.0  &  III  \\ 
G341.997$-$00.007  &  341.997  &  $-$0.007  &  13.7  &  III  \\ 
G342.065+01.000  &  342.065  &  1.000  &  33.2  &  II  \\ 
G342.427+00.472  &  342.427  &  0.472  &  18.8  &  II  \\ 
G342.556+00.615  &  342.556  &  0.615  &  7.5  &  I  \\ 
G342.731+00.172  &  342.731  &  0.172  &  2.7  &  II  \\ 
G343.326$-$00.335  &  343.326  &  $-$0.335  &  5.0  &  II  \\ 
G343.591$-$00.071  &  343.591  &  $-$0.071  &  12.6  &  II  \\ 
G344.064$-$00.013  &  344.064  &  $-$0.013  &  6.6  &  III  \\ 
G345.167$-$00.172  &  345.167  &  $-$0.172  &  6.4  &  II  \\ 
G345.308$-$01.077  &  345.308  &  $-$1.077  &  3.4  &  II  \\ 
G345.532$-$00.001  &  345.532  &  $-$0.001  &  4.4  &  I  \\ 
G345.821+01.199  &  345.821  &  1.199  &  10.5  &  II  \\ 
G345.899+00.169  &  345.899  &  0.169  &  4.3  &  I  \\ 
G346.072+00.040  &  346.072  &  0.040  &  2.3  &  II  \\ 
G346.365+00.305  &  346.365  &  0.305  &  3.8  &  III  \\ 
G346.821$-$00.981  &  346.821  &  $-$0.981  &  6.0  &  III  \\ 
G346.849+00.353  &  346.849  &  0.353  &  3.5  &  II  \\ 
G346.907+00.323  &  346.907  &  0.323  &  1.0  &  II  \\ 
G347.362$-$00.352  &  347.362  &  $-$0.352  &  15.8  &  I  \\ 
G347.380+00.473  &  347.380  &  0.473  &  19.0  &  II  \\ 
G348.160$-$00.211  &  348.160  &  $-$0.211  &  12.0  &  II  \\ 
G348.324$-$00.735  &  348.324  &  $-$0.735  &  4.7  &  I  \\ 
G348.374+00.153  &  348.374  &  0.153  &  9.0  &  I  \\ 
G348.586+00.249  &  348.586  &  0.249  &  7.3  &  I  \\ 
G350.299$-$00.086  &  350.299  &  $-$0.086  &  9.7  &  I  \\ 
G350.404$-$01.013  &  350.404  &  $-$1.013  &  2.6  &  II  \\ 
G350.471+00.088  &  350.471  &  0.088  &  1.3  &  II  \\ 
G350.584$-$00.909  &  350.584  &  $-$0.909  &  2.8  &  II  \\ 
G350.937$-$00.338  &  350.937  &  $-$0.338  &  18.5  &  II  \\ 
G351.361$-$00.049  &  351.361  &  $-$0.049  &  9.9  &  I  \\ 
G351.658+00.093  &  351.658  &  0.093  &  5.9  &  I  \\ 
G351.790$-$00.239  &  351.790  &  $-$0.239  &  7.7  &  II  \\ 
G352.220$-$00.458  &  352.220  &  $-$0.458  &  18.1  &  II  \\ 
G352.267$-$00.062  &  352.267  &  $-$0.062  &  7.4  &  I  \\ 
G352.589$-$00.299  &  352.589  &  $-$0.299  &  7.4  &  II  \\ 
G352.759$-$00.340  &  352.759  &  $-$0.340  &  5.4  &  I  \\ 
G354.353$-$00.018  &  354.353  &  $-$0.018  &  2.3  &  II  \\ 
G354.353$-$00.018  &  354.353  &  $-$0.018  &  2.3  &  II  \\ 
G354.362+00.285  &  354.362  &  0.285  &  9.9  &  II  \\ 
G354.931$-$00.241  &  354.931  &  $-$0.241  &  2.9  &  II  \\ 
G355.406+00.513  &  355.406  &  0.513  &  3.5  &  II  \\ 
G355.464$-$00.253  &  355.464  &  $-$0.253  &  3.0  &  II  \\ 
G355.854$-$00.084  &  355.854  &  $-$0.084  &  1.0  &  III  \\ 
G355.919$-$00.206  &  355.919  &  $-$0.206  &  2.0  &  I  \\ 
G356.621+00.125  &  356.621  &  0.125  &  4.2  &  III  \\ 
G356.662$-$00.297  &  356.662  &  $-$0.297  &  12.8  &  II  \\ 
G356.717+00.139  &  356.717  &  0.139  &  5.1  &  II  \\ 
G356.795$-$00.073  &  356.795  &  $-$0.073  &  0.8  &  II  \\ 
G357.083+00.318  &  357.083  &  0.318  &  3.2  &  I  \\ 
G357.109$-$00.187  &  357.109  &  $-$0.187  &  9.0  &  I  \\ 
G357.109+00.055  &  357.109  &  0.055  &  5.2  &  I  \\ 
G357.528$-$00.455  &  357.528  &  $-$0.455  &  11.2  &  III  \\ 
G357.976$-$00.145  &  357.976  &  $-$0.145  &  0.9  &  II  \\ 
G358.280$-$00.850  &  358.280  &  $-$0.850  &  14.0  &  I  \\ 
G358.282+00.069  &  358.282  &  0.069  &  2.5  &  II  \\ 
G358.398+00.090  &  358.398  &  0.090  &  2.6  &  I  \\ 
G001.964$-$00.573  &  1.964  &  $-$0.573  &  1.2  &  III  \\ 
G002.684+00.032  &  2.684  &  0.032  &  12.2  &  III  \\ 
G003.270+00.128  &  3.270  &  0.128  &  5.6  &  II  \\ 
G003.656$-$00.124  &  3.656  &  $-$0.124  &  5.5  &  III  \\ 
G004.215+00.351  &  4.215  &  0.351  &  0.7  &  II  \\ 
G005.522+01.080  &  5.522  &  1.080  &  7.6  &  II  \\ 
G005.996$-$00.577  &  5.996  &  $-$0.577  &  1.3  &  II  \\ 
G006.234$-$00.784  &  6.234  &  $-$0.784  &  1.2  &  III  \\ 
G006.249$-$00.757  &  6.249  &  $-$0.757  &  0.9  &  III  \\ 
G006.609$-$00.608  &  6.609  &  $-$0.608  &  11.7  &  II  \\ 
G007.374$-$00.149  &  7.374  &  $-$0.149  &  6.0  &  III  \\ 
G007.377+00.716  &  7.377  &  0.716  &  12.3  &  I  \\ 
G008.240+00.472  &  8.240  &  0.472  &  18.4  &  III  \\ 
G008.482+00.202  &  8.482  &  0.202  &  11.4  &  III  \\ 
G009.211+00.063  &  9.211  &  0.063  &  4.7  &  II  \\ 
G009.367+00.444  &  9.367  &  0.444  &  12.9  &  II  \\ 
G009.811$-$00.083  &  9.811  &  $-$0.083  &  7.8  &  II  \\ 
G010.026$-$00.262  &  10.026  &  $-$0.262  &  1.9  &  II  \\ 
G010.670$-$00.376  &  10.670  &  $-$0.376  &  2.3  &  I  \\ 
G012.014$-$00.576  &  12.014  &  $-$0.576  &  6.4  &  I  \\ 
G012.864+00.010  &  12.864  &  0.010  &  9.8  &  III  \\ 
G013.155$-$00.797  &  13.155  &  $-$0.797  &  4.8  &  II  \\ 
G013.313$-$00.695  &  13.313  &  $-$0.695  &  6.6  &  III  \\ 
G013.593$-$00.377  &  13.593  &  $-$0.377  &  6.8  &  II  \\ 
G014.223$-$00.174  &  14.223  &  $-$0.174  &  13.2  &  III  \\ 
G017.218$-$00.077  &  17.218  &  $-$0.077  &  4.9  &  II  \\ 
G021.466+00.135  &  21.466  &  0.135  &  6.2  &  III  \\ 
G023.293$-$00.563  &  23.293  &  $-$0.563  &  10.2  &  II  \\ 
G026.902+00.240  &  26.902  &  0.240  &  5.1  &  III  \\ 
G032.582+00.779  &  32.582  &  0.779  &  2.3  &  II  \\ 
G035.997+00.957  &  35.997  &  0.957  &  11.2  &  III  \\ 
G036.245+00.350  &  36.245  &  0.350  &  10.5  &  II  \\ 
G036.858+00.083  &  36.858  &  0.083  &  5.6  &  II  \\ 
G037.455$-$00.282  &  37.455  &  $-$0.282  &  4.7  &  III  \\ 
G037.518$-$00.631  &  37.518  &  $-$0.631  &  3.5  &  II  \\ 
G037.693+01.814  &  37.693  &  1.814  &  58.0  &  III  \\ 
G037.877+00.593  &  37.877  &  0.593  &  3.1  &  II  \\ 
G038.214$-$00.202  &  38.214  &  $-$0.202  &  1.0  &  II  \\ 
G038.815$-$00.136  &  38.815  &  $-$0.136  &  4.3  &  I  \\ 
G039.038+00.748  &  39.038  &  0.748  &  11.7  &  I  \\ 
G039.516+00.501  &  39.516  &  0.501  &  6.1  &  I  \\ 
G040.341+00.277  &  40.341  &  0.277  &  1.6  &  II  \\ 
G040.449+00.540  &  40.449  &  0.540  &  9.1  &  I  \\ 
G040.866+00.155  &  40.866  &  0.155  &  10.6  &  II  \\ 
G041.699+00.975  &  41.699  &  0.975  &  3.0  &  II  \\ 
G042.186$-$00.802  &  42.186  &  $-$0.802  &  2.0  &  II  \\ 
G042.506$-$00.445  &  42.506  &  $-$0.445  &  1.7  &  III  \\ 
G042.665$-$00.634  &  42.665  &  $-$0.634  &  1.8  &  III  \\ 
G042.954+00.074  &  42.954  &  0.074  &  4.4  &  I  \\ 
G043.047$-$00.067  &  43.047  &  $-$0.067  &  5.4  &  I  \\ 
G046.304+00.352  &  46.304  &  0.352  &  11.2  &  III  \\ 
G046.577+00.221  &  46.577  &  0.221  &  9.0  &  III  \\ 
G043.942+01.442  &  43.942  &  1.442  &  8.1  &  I  \\ 
G044.106+00.595  &  44.106  &  0.595  &  24.2  &  III  \\ 
G044.339+01.273  &  44.339  &  1.273  &  2.5  &  I  \\ 
G046.420$-$00.169  &  46.420  &  $-$0.169  &  9.3  &  I  \\ 
G046.910+00.114  &  46.910  &  0.114  &  7.3  &  I  \\ 
G050.489$-$00.410  &  50.489  &  $-$0.410  &  8.9  &  II  \\ 
G051.746+01.384  &  51.746  &  1.384  &  8.6  &  II  \\ 
G055.226+00.477  &  55.226  &  0.477  &  16.4  &  III  \\ 
G056.116+00.294  &  56.116  &  0.294  &  2.9  &  II  \\ 
G057.423+00.864  &  57.423  &  0.864  &  11.9  &  I  \\ 
G057.675+00.676  &  57.675  &  0.676  &  11.9  &  I  \\ 
G057.868+00.221  &  57.868  &  0.221  &  8.4  &  I  \\ 
G057.884+00.543  &  57.884  &  0.543  &  11.9  &  I  \\ 
G058.285+00.307  &  58.285  &  0.307  &  1.7  &  II  \\ 

\end{xtabular}
\tablefoot{\tablefoottext{a}{``I'' is most reliable and have clear SMGPS emission that is not confused with that of other SNRs or \hii\ regions, with a morphology that is similar to that of known SNRs; ``II'' is less reliable and are either in confused regions with other SNRs or \hii\ regions or have a morphology that is slightly ambiguous; ``III'' are the least reliable and are either badly confused with other SNRs or \hii\ regions, have spatially coincident radio and MIR emission that may indicate that at least some of the radio emission is thermal, or are incomplete shells.}}
\endgroup
\vskip 15pt

\subsection{Unusual Sources\label{sec:odd}}
\begin{figure*}
    \centering
    \includegraphics[width=3.6in]{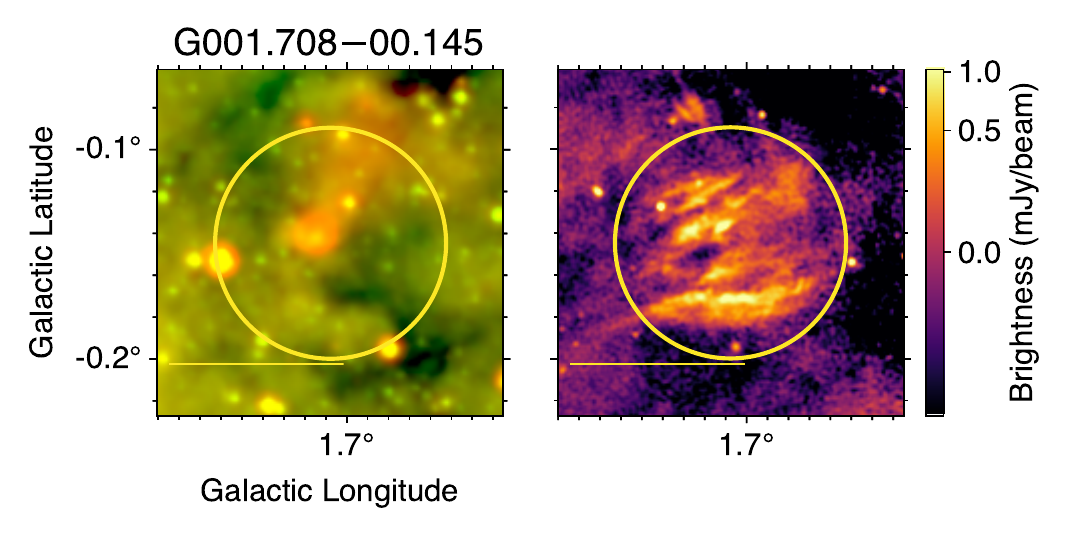}
    \includegraphics[width=3.6in]{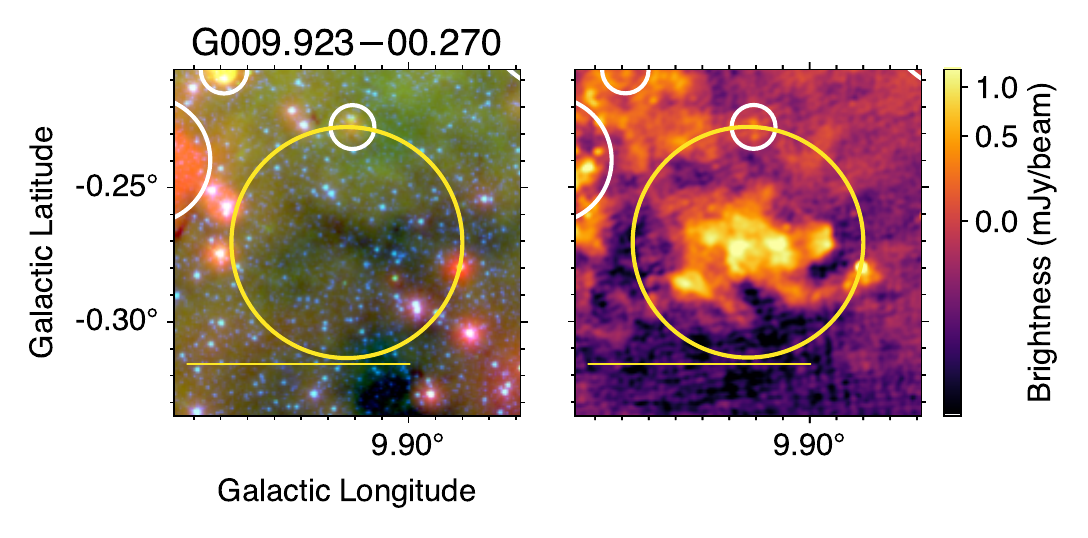}
    \caption{Example unusual sources of SMGPS emission. MIR data is in the left panels (Spitzer or WISE) and SMGPS 1.3\,\ghz\ data are on the right.  The symbols have the same meaning as in Figure~\ref{fig:overview}, with the unusual sources being represented by yellow circles at the centers of all panels.  Yellow scale bars at the lower left of each panel are $5\arcmin$ long.}
    \label{fig:odd}
\end{figure*}

We identify 49
unusual sources of radio continuum emission. These sources do not have the expected radio morphology for SNRs (or \hii\ regions) but they do lack MIR emission.  Included in this list is known SNR G298.5$-$00.3 (Figure~\ref{fig:G22_bad}).  These objects may be SNRs, but because they lack the characteristic SNR radio morphologies, we do not include them in the SNR candidates list. They are worthy of future study.
We list the position and circular size of these regions in Table~\ref{tab:odd} and show examples in Figure~\ref{fig:odd}.

\vskip 15pt
\begingroup\small
\topcaption{Unusual SMGPS Sources\label{tab:odd}}
\tablefirsthead{
\hline\hline
Name & GLong & GLat & Radius\\
 &  deg. &  deg. &  arcmin. \\
\hline}
\tablehead{%
\multicolumn{4}{@{}l}{Table \thetable, continued.}\\
\hline\hline
Name & GLong & GLat & Radius\\
 &  deg. &  deg. &  arcmin.\\
\hline}
\tabletail{\hline}
\tablelasttail{\hline}
\begin{xtabular}{lrrrl}
G274.523$-$01.021  &  274.523  &  $-$1.021  &  5.6  \\ 
G279.423$-$01.431  &  279.423  &  $-$1.431  &  4.2  \\ 
G282.022$-$00.782  &  282.022  &  $-$0.782  &  2.2  \\ 
G284.217+00.400  &  284.217  &  0.400  &  8.6  \\ 
G289.104$-$00.563  &  289.104  &  $-$0.563  &  2.9  \\ 
G293.650$-$00.229  &  293.650  &  $-$0.229  &  8.5  \\ 
G298.516$-$00.352  &  298.516  &  $-$0.352  &  4.2  \\ 
G301.980$-$00.316  &  301.980  &  $-$0.316  &  1.2  \\ 
G310.863+00.011  &  310.863  &  0.011  &  1.1  \\ 
G314.085$-$00.486  &  314.085  &  $-$0.486  &  6.0  \\ 
G315.487$-$00.790  &  315.487  &  $-$0.790  &  0.6  \\ 
G319.685+00.121  &  319.685  &  0.121  &  2.1  \\ 
G322.113+00.879  &  322.113  &  0.879  &  6.7  \\ 
G324.067$-$00.581  &  324.067  &  $-$0.581  &  4.3  \\ 
G324.633$-$00.410  &  324.633  &  $-$0.410  &  1.8  \\ 
G325.709+00.959  &  325.709  &  0.959  &  10.5  \\ 
G329.877$-$00.458  &  329.877  &  $-$0.458  &  0.7  \\ 
G332.176+01.426  &  332.176  &  1.426  &  1.0  \\ 
G332.228+00.198  &  332.228  &  0.198  &  1.1  \\ 
G336.953$-$00.046  &  336.953  &  $-$0.046  &  2.1  \\ 
G337.412+00.077  &  337.412  &  0.077  &  2.9  \\ 
G339.388+00.700  &  339.388  &  0.700  &  1.7  \\ 
G341.503$-$00.424  &  341.503  &  $-$0.424  &  0.9  \\ 
G344.833$-$00.028  &  344.833  &  $-$0.028  &  1.6  \\ 
G345.685+00.193  &  345.685  &  0.193  &  1.0  \\ 
G345.884$-$00.172  &  345.884  &  $-$0.172  &  2.1  \\ 
G346.169$-$00.419  &  346.169  &  $-$0.419  &  5.0  \\ 
G346.366+00.105  &  346.366  &  0.105  &  3.1  \\ 
G346.378$-$00.024  &  346.378  &  $-$0.024  &  2.1  \\ 
G351.058+00.022  &  351.058  &  0.022  &  2.2  \\ 
G354.136$-$00.185  &  354.136  &  $-$0.185  &  2.4  \\ 
G354.191+00.620  &  354.191  &  0.620  &  7.9  \\ 
G354.299$-$00.066  &  354.299  &  $-$0.066  &  1.3  \\ 
G355.946+00.192  &  355.946  &  0.192  &  3.3  \\ 
G356.489$-$00.007  &  356.489  &  $-$0.007  &  2.8  \\ 
G001.708$-$00.145  &  1.708  &  $-$0.145  &  3.3  \\ 
G001.763+00.057  &  1.763  &  0.057  &  2.4  \\ 
G002.361$-$00.134  &  2.361  &  $-$0.134  &  9.5  \\ 
G002.805+00.131  &  2.805  &  0.131  &  3.1  \\ 
G005.489$-$00.350  &  5.489  &  $-$0.350  &  5.5  \\ 
G006.195+00.651  &  6.195  &  0.651  &  2.7  \\ 
G009.923$-$00.270  &  9.923  &  $-$0.270  &  2.6  \\ 
G013.379+00.138  &  13.379  &  0.138  &  5.4  \\ 
G017.307+00.095  &  17.307  &  0.095  &  2.7  \\ 
G018.513+00.077  &  18.513  &  0.077  &  7.9  \\ 
G019.175+00.653  &  19.175  &  0.653  &  3.7  \\ 
G019.225+00.823  &  19.225  &  0.823  &  3.0  \\ 
G043.335$-$00.252  &  43.335  &  $-$0.252  &  1.1  \\ 
G051.564+00.095  &  51.564  &  0.095  &  0.7  \\ 

\end{xtabular}
\endgroup
\vskip 15pt

\subsection{Comparison Between the Galactic SNR, SNR Candidate, and HII Region Populations}
\begin{figure*}
    \centering
    \includegraphics[width=\textwidth]{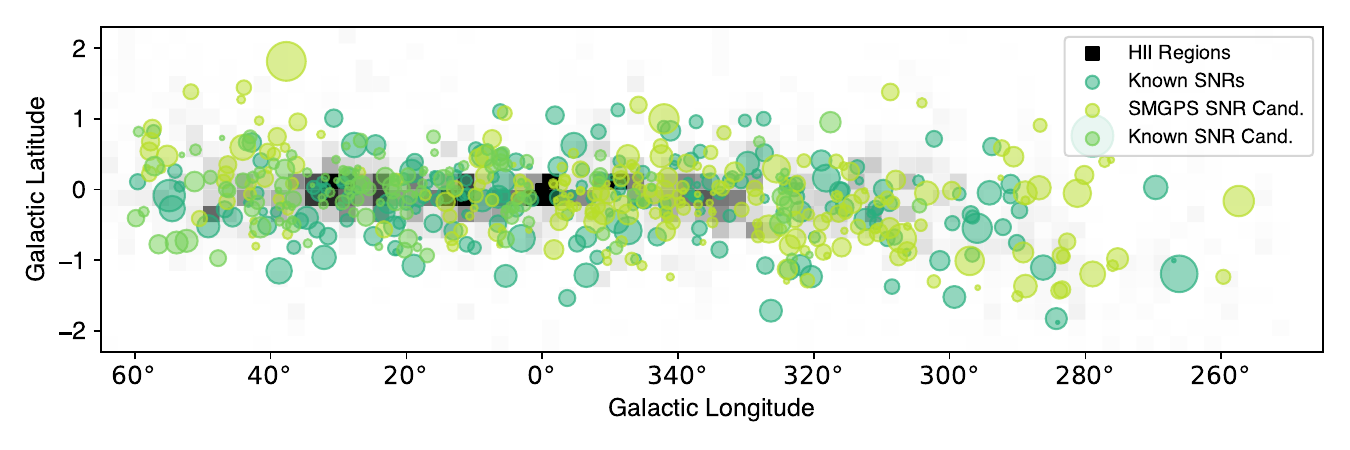}
    \caption{Galactic distribution of the previously-known SNRs (blue-green, from both the G22 and SNRCat samples),
    SMGPS SNR candidates (light green filled), and previously-known SNR candidates (green open).  The circles are representative of the SNR sizes.  Only sources confirmed here are shown.  The background is a two-dimensional histogram of the \hii\ region density from the WISE Catalog, with higher densities indicated by darker shades.}
    \label{fig:pospos}
\end{figure*}

The areal distributions of SNRs and SNR candidates, shown in Figure~\ref{fig:pospos}, are visually similar to each other.  To assess whether the various samples are statistically distinct, we use a Kolmogorov-Smirnov (K-S) test, adopting $p=0.001$ as the discriminating value such that statistically distinct samples will have $p<0.001$.

\begin{figure*}
    \centering
    \includegraphics[width=3.6in]{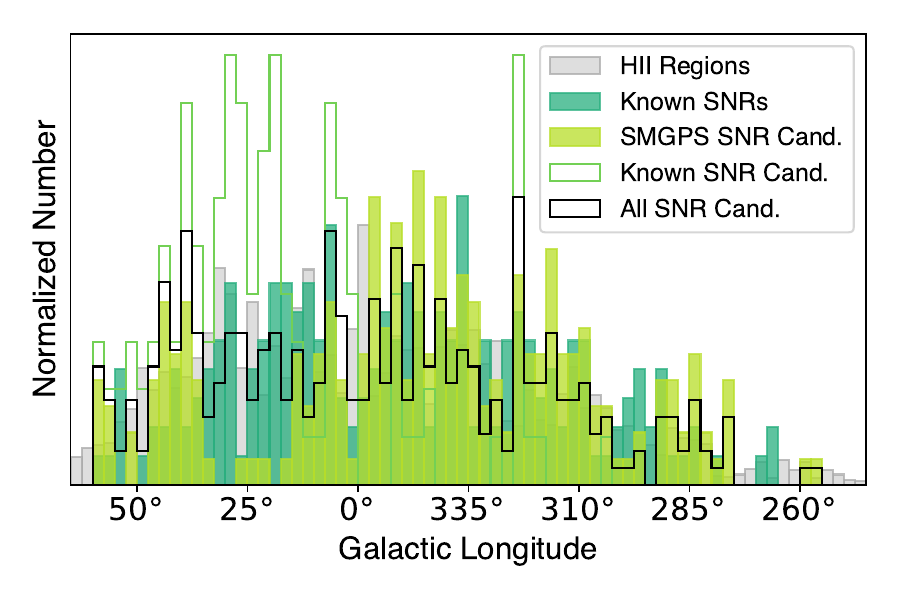}
    \includegraphics[width=3.6in]{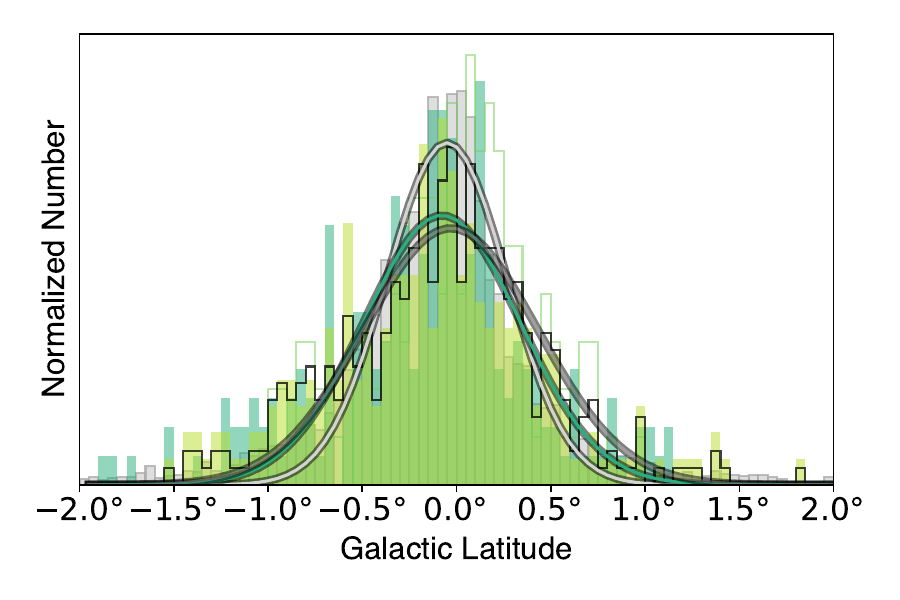}
    \caption{Galactic longitude (left) and latitude (right) distributions for
    WISE Catalog \hii\ regions (grey),
    known SNRs (blue-green, from both the G22 and SNRCat samples),
    SMGPS SNR candidates (light green filled), previously-known SNR candidates (green open), and all SNR candidates (SMGPS-discovered and previously-known combined; black open). Only sources confirmed here are shown. Curves in the right panel show least-squares fits to the binned distributions - the grey curve for \hii\ regions, the blue-green curve for known SNRs, and the black curve for all SNR candidates.}
    \label{fig:glong_glat}
\end{figure*}

We show the Galactic longitude distributions in the left panel of Figure~\ref{fig:glong_glat}.
Because of the increased sensitivity of the SMGPS data compared to that previously available, there are far more newly-identified SNR candidates in the fourth versus the first Galactic quadrant (72 versus 28\%).  
SNRs are more prevalent relative to \hii\ regions between $300\degree <\ell < 250\degree$. The SNR candidate longitude distribution (including all SMGPS SNR candidates and those previously identified) is not statistically distinct from the known SNR distribution ($p=0.017$).  

We show the Galactic latitude distributions in the right panel of Figure~\ref{fig:glong_glat}.
The SNR candidate latitude distribution (including all SMGPS SNR candidates and those previously identified) is not statistically distinct from the known SNR distribution ($p=0.26$).  Least-squares fits to the binned Galactic latitude distributions of known and candidate SNRs (including both previously-known and SMGPS discovered) have full width at half maximum (FWHM) values of 0.97\degree, and 1.08\degree, respectively.  The \hii\ region distribution has a higher percentage of sources near $b=0\degree$; its FWHM fit value is 0.75\degree.  Measured from the standard deviation $\sigma$, however, all three samples have similar FWHM values (FWHM$ = 2.354\sigma$).  We find
the FWHM for the samples of known SNRs, candidate SNRs, and \hii\ regions is 1.29\degree, 1.25\degree, and 1.22\degree, respectively. 

\begin{figure}
    \centering
    \includegraphics[width=3.6in]{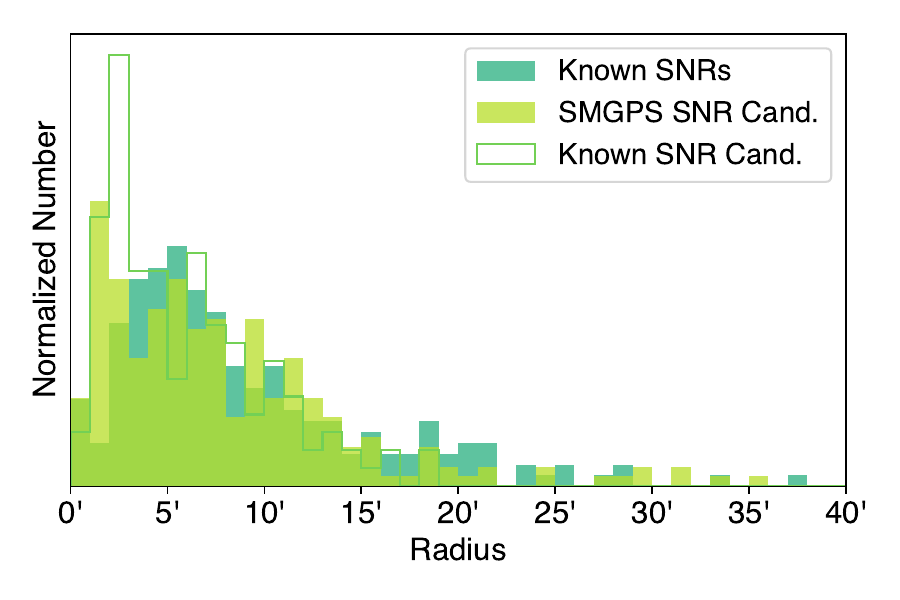}
    \caption{Angular radius distributions for and known SNRs (red, from both the G22 and SNRCat samples) SNRs, SMGPS SNR candidates (light green filled), and previously-known SNR candidates (green open). Only sources detected in SMGPS images are shown.}
    \label{fig:radius}
\end{figure}

As shown in Fig.~\ref{fig:radius}, the known and candidate SNR angular radius distributions are similar, but the SNR candidate distributions have a higher percentage of sources at lower values.  A K-S test indicates that they are marginally statistically distinct ($p=0.001$).  Given the higher angular resolution of recent radio surveys whose data has been used to identify SNR candidates (e.g., THOR, GLOSTAR, and the SMGPS), this difference is unsurprising.


\subsection{Implications for the Galactic SNR population}
If confirmed (i.e. with radio spectral index or polarization measurements, or with optical or x-ray searches), the 237
new SNR candidates would approximately double the
known SNR population in the survey zone and would
increase the total Galactic SNR population by $\sim 80\%$ (using the G22 number).
Even if all SNR candidates are confirmed, however, there would probably be hundreds to thousands of Galactic SNRs that are yet to be identified (see Section~\ref{sec:intro}).  Future radio continuum surveys will hopefully find them.  


\section{Summary}
Using 1.3\,\ghz\ continuum data from the MeerKAT Galactic Plane survey we cataloged
237
new Galactic SNR candidates.  Our method identifies
diffuse radio continuum emission regions lacking the MIR emission
counterparts that would be seen for \hii\ regions and planetary nebulae.  All
candidates lack MIR emission from known \hii\ regions, as cataloged in
the WISE Catalog of Galactic \hii\ Regions.  The detected candidates
follow a similar areal distribution and have similar radii compared to the previously-identified
sample.

We also detect radio continuum emission and confirm the expected morphology for 201 known SNRs: 184
from
\citet{green22} and 30
from \citet{ferrand12}.  We identify two SNRs from \citet{green22} and 21
from \citet{ferrand12} that we believe should be reclassified.

If our candidates prove to be true SNRs, they would approximately double
the Galactic SNR population in the surveyed region and increase the known SNR population by $\sim\!80\%$.   Although this is still fewer than expected for a Milky Way-type galaxy, the discrepancy is far less severe than it was a few years ago.  There are now $>300$ Galactic SNR candidates, indicating the need for followup observations.

\begin{acknowledgements}
{We thank the anonymous referee for constructive comments that increased the clarity and readability of this work.}
The MeerKAT telescope is operated by the South African Radio Astronomy Observatory, which is a facility of the National Research Foundation, an agency of the Department of Science and Innovation. This research has made use of the SIMBAD database,
operated at CDS, Strasbourg, France. LDA thanks Cara Gillotti for her invaluable help drawing green circles.  LDA and TF acknowledge support from NSF ASTR \#2307176.
\nraoblurb
\end{acknowledgements}

\bibliographystyle{aa.bst}
\bibliography{ref.bib}

\begin{thebibliography}{56}
\expandafter\ifx\csname natexlab\endcsname\relax\def\natexlab#1{#1}\fi

\bibitem[{{Anderson} {et~al.}(2014){Anderson}, {Bania}, {Balser}, {Cunningham}, {Wenger}, {Johnstone}, \& {Armentrout}}]{anderson14}
{Anderson}, L.~D., {Bania}, T.~M., {Balser}, D.~S., {et~al.} 2014, \apjs, 212, 1

\bibitem[{{Anderson} {et~al.}(2011){Anderson}, {Bania}, {Balser}, \& {Rood}}]{anderson11}
{Anderson}, L.~D., {Bania}, T.~M., {Balser}, D.~S., \& {Rood}, R.~T. 2011, \apjs, 194, 32

\bibitem[{{Anderson} {et~al.}(2017){Anderson}, {Wang}, {Bihr}, {Rugel}, {Beuther}, {Bigiel}, {Churchwell}, {Glover}, {Goodman}, {Henning}, {Heyer}, {Klessen}, {Linz}, {Longmore}, {Menten}, {Ott}, {Roy}, {Soler}, {Stil}, \& {Urquhart}}]{anderson17}
{Anderson}, L.~D., {Wang}, Y., {Bihr}, S., {et~al.} 2017, \aap, 605, A58

\bibitem[{{Armentrout} {et~al.}(2021){Armentrout}, {Anderson}, {Wenger}, {Balser}, \& {Bania}}]{armentrout21}
{Armentrout}, W.~P., {Anderson}, L.~D., {Wenger}, T.~V., {Balser}, D.~S., \& {Bania}, T.~M. 2021, \apjs, 253, 23

\bibitem[{{Ball} {et~al.}(2023){Ball}, {Kothes}, {Rosolowsky}, {West}, {Becker}, {Filipovi{\'c}}, {Gaensler}, {Hopkins}, {Koribalski}, {Landecker}, {Leahy}, {Marvil}, {Sun}, {Bufano}, {Carretti}, {Ingallinera}, {Van Eck}, \& {Willis}}]{ball23}
{Ball}, B.~D., {Kothes}, R., {Rosolowsky}, E., {et~al.} 2023, \mnras, 524, 1396

\bibitem[{{Bamba} {et~al.}(2003){Bamba}, {Ueno}, {Koyama}, \& {Yamauchi}}]{bamba03}
{Bamba}, A., {Ueno}, M., {Koyama}, K., \& {Yamauchi}, S. 2003, \apj, 589, 253

\bibitem[{{Benjamin} {et~al.}(2003){Benjamin}, {Churchwell}, {Babler}, {Bania}, {Clemens}, {Cohen}, {Dickey}, {Indebetouw}, {Jackson}, {Kobulnicky}, {Lazarian}, {Marston}, {Mathis}, {Meade}, {Seager}, {Stolovy}, {Watson}, {Whitney}, {Wolff}, \& {Wolfire}}]{benjamin03}
{Benjamin}, R.~A., {Churchwell}, E., {Babler}, B.~L., {et~al.} 2003, \pasp, 115, 953

\bibitem[{{Beuther} {et~al.}(2016){Beuther}, {Bihr}, {Rugel}, {Johnston}, {Wang}, {Walter}, {Brunthaler}, {Walsh}, {Ott}, {Stil}, {Henning}, {Schierhuber}, {Kainulainen}, {Heyer}, {Goldsmith}, {Anderson}, {Longmore}, {Klessen}, {Glover}, {Urquhart}, {Plume}, {Ragan}, {Schneider}, {McClure-Griffiths}, {Menten}, {Smith}, {Roy}, {Shanahan}, {Nguyen-Luong}, \& {Bigiel}}]{beuther16}
{Beuther}, H., {Bihr}, S., {Rugel}, M., {et~al.} 2016, \aap, 595, A32

\bibitem[{{Bietenholz} \& {Bartel}(2008)}]{bietenholz08}
{Bietenholz}, M.~F. \& {Bartel}, N. 2008, \mnras, 386, 1411

\bibitem[{{Bietenholz} {et~al.}(2015){Bietenholz}, {Yuan}, {Buehler}, {Lobanov}, \& {Blandford}}]{bietenholz15b}
{Bietenholz}, M.~F., {Yuan}, Y., {Buehler}, R., {Lobanov}, A.~P., \& {Blandford}, R. 2015, \mnras, 446, 205

\bibitem[{{Brogan} {et~al.}(2006){Brogan}, {Gelfand}, {Gaensler}, {Kassim}, \& {Lazio}}]{brogan06}
{Brogan}, C.~L., {Gelfand}, J.~D., {Gaensler}, B.~M., {Kassim}, N.~E., \& {Lazio}, T.~J.~W. 2006, \apjl, 639, L25

\bibitem[{Camilo {et~al.}(2018)Camilo, Scholz, Serylak, \& et~al.}]{Camilo2018}
Camilo, F., Scholz, P., Serylak, M., \& et~al. 2018, ApJ, 856, 180

\bibitem[{{Carey} {et~al.}(2009){Carey}, {Noriega-Crespo}, {Mizuno}, {Shenoy}, {Paladini}, {Kraemer}, {Price}, {Flagey}, {Ryan}, {Ingalls}, {Kuchar}, {Pinheiro Gon{\c c}alves}, {Indebetouw}, {Billot}, {Marleau}, {Padgett}, {Rebull}, {Bressert}, {Ali}, {Molinari}, {Martin}, {Berriman}, {Boulanger}, {Latter}, {Miville-Deschenes}, {Shipman}, \& {Testi}}]{carey09}
{Carey}, S.~J., {Noriega-Crespo}, A., {Mizuno}, D.~R., {et~al.} 2009, \pasp, 121, 76

\bibitem[{{Churchwell} {et~al.}(2009){Churchwell}, {Babler}, {Meade}, {Whitney}, {Benjamin}, {Indebetouw}, {Cyganowski}, {Robitaille}, {Povich}, {Watson}, \& {Bracker}}]{churchwell09}
{Churchwell}, E., {Babler}, B.~L., {Meade}, M.~R., {et~al.} 2009, \pasp, 121, 213

\bibitem[{{Cohen} \& {Green}(2001)}]{cohen01}
{Cohen}, M. \& {Green}, A.~J. 2001, \mnras, 325, 531

\bibitem[{{Cotton} {et~al.}(2024){Cotton}, {Kothes}, {Camilo}, {Chandra}, {Buchner}, \& {Nyamai}}]{cotton24}
{Cotton}, W.~D., {Kothes}, R., {Camilo}, F., {et~al.} 2024, \apjs, 270, 21

\bibitem[{{de Avillez} \& {Breitschwerdt}(2005)}]{deavillez05}
{de Avillez}, M.~A. \& {Breitschwerdt}, D. 2005, \aap, 436, 585

\bibitem[{{Dokara} {et~al.}(2021){Dokara}, {Brunthaler}, {Menten}, {Dzib}, {Reich}, {Cotton}, {Anderson}, {Chen}, {Gong}, {Medina}, {Ortiz-Le{\'o}n}, {Rugel}, {Urquhart}, {Wyrowski}, {Yang}, {Beuther}, {Billington}, {Csengeri}, {Carrasco-Gonz{\'a}lez}, \& {Roy}}]{dokara21}
{Dokara}, R., {Brunthaler}, A., {Menten}, K.~M., {et~al.} 2021, arXiv e-prints, arXiv:2103.06267

\bibitem[{{Dokara} {et~al.}(2018){Dokara}, {Roy}, {Beuther}, {Anderson}, {Rugel}, {Stil}, {Wang}, {Soler}, \& {Shanahan}}]{dokara18}
{Dokara}, R., {Roy}, N., {Beuther}, H., {et~al.} 2018, \apj, 866, 61

\bibitem[{{Driessen} {et~al.}(2018){Driessen}, {Dom{\v{c}}ek}, {Vink}, {Hessels}, {Arias}, \& {Gelfand }}]{driessen18}
{Driessen}, L.~N., {Dom{\v{c}}ek}, V., {Vink}, J., {et~al.} 2018, \apj, 860, 133

\bibitem[{{Dubner} \& {Giacani}(2015)}]{dubner15}
{Dubner}, G. \& {Giacani}, E. 2015, \aapr, 23, 3

\bibitem[{{Faucher-Gigu{\`e}re} {et~al.}(2013){Faucher-Gigu{\`e}re}, {Quataert}, \& {Hopkins}}]{faucher13}
{Faucher-Gigu{\`e}re}, C.-A., {Quataert}, E., \& {Hopkins}, P.~F. 2013, \mnras, 433, 1970

\bibitem[{{Ferrand} \& {Safi-Harb}(2012)}]{ferrand12}
{Ferrand}, G. \& {Safi-Harb}, S. 2012, Advances in Space Research, 49, 1313

\bibitem[{{Gaensler} {et~al.}(2001){Gaensler}, {Slane}, {Gotthelf}, \& {Vasisht}}]{gaensler01}
{Gaensler}, B.~M., {Slane}, P.~O., {Gotthelf}, E.~V., \& {Vasisht}, G. 2001, \apj, 559, 963

\bibitem[{{Girichidis} {et~al.}(2016){Girichidis}, {Walch}, {Naab}, {Gatto}, {W{\"u}nsch}, {Glover}, {Klessen}, {Clark}, {Peters}, {Derigs}, \& {Baczynski}}]{girichidis16}
{Girichidis}, P., {Walch}, S., {Naab}, T., {et~al.} 2016, \mnras, 456, 3432

\bibitem[{{Goedhart} {et~al.}(2024){Goedhart}, {Cotton}, {Camilo}, {Thompson}, {Umana}, {Bietenholz}, {Woudt}, {Anderson}, {Bordiu}, {Buckley}, {Buemi}, {Bufano}, {Cavallaro}, {Chen}, {Chibueze}, {Egbo}, {Frank}, {Hoare}, {Ingallinera}, {Irabor}, {Kraan-Korteweg}, {Kurapati}, {Leto}, {Loru}, {Mutale}, {Obonyo}, {Plavin}, {Rajohnson}, {Rigby}, {Riggi}, {Seidu}, {Serra}, {Smart}, {Stappers}, {Steyn}, {Surnis}, {Trigilio}, {Williams}, {Abbott}, {Adam}, {Asad}, {Baloyi}, {Bauermeister}, {Bennet}, {Bester}, {Botha}, {Brederode}, {Buchner}, {Burger}, {Cheetham}, {Cloete}, {de Villiers}, {de Villiers}, {du Toit}, {Esterhuyse}, {Fanaroff}, {Fourie}, {Gamatham}, {Gatsi}, {Geyer}, {Gouws}, {Gumede}, {Heywood}, {Hokwana}, {Hoosen}, {Horn}, {Horrell}, {Hugo}, {Isaacson}, {J{\'o}zsa}, {Jonas}, {Jordaan}, {Joubert}, {Julie}, {Kapp}, {Kriek}, {Kriel}, {Krishnan}, {Kusel}, {Legodi}, {Lehmensiek}, {Lord}, {Macfarlane}, {Magnus}, {Magozore}, {Main}, {Malan}, {Manley}, {Marais}, {Maree}, {Martens}, {Maruping}, {McAlpine},
  {Merry}, {Mgodeli}, {Millenaar}, {Mokone}, {Monama}, {New}, {Ngcebetsha}, {Ngoasheng}, {Nicolson}, {Ockards}, {Oozeer}, {Passmoor}, {Patel}, {Peens-Hough}, {Perkins}, {Ramaila}, {Ratcliffe}, {Renil}, {Richter}, {Salie}, {Sambu}, {Schollar}, {Schwardt}, {Schwartz}, {Serylak}, {Siebrits}, {Sirothia}, {Slabber}, {Smirnov}, {Tiplady}, {van Balla}, {van der Byl}, {Van Tonder}, {Venter}, {Venter}, {Welz}, \& {Williams}}]{goedhart24}
{Goedhart}, S., {Cotton}, W.~D., {Camilo}, F., {et~al.} 2024, \mnras, 531, 649

\bibitem[{Gonçalves {et~al.}(2011)Gonçalves, Noriega-Crespo, Paladini, Martin, \& Carey}]{goncalves11}
Gonçalves, D.~P., Noriega-Crespo, A., Paladini, R., Martin, P.~G., \& Carey, S.~J. 2011, The Astronomical Journal, 142, 47

\bibitem[{{Green} {et~al.}(2014){Green}, {Reeves}, \& {Murphy}}]{green14b}
{Green}, A.~J., {Reeves}, S.~N., \& {Murphy}, T. 2014, \pasa, 31, e042

\bibitem[{{Green}(2004)}]{green04}
{Green}, D.~A. 2004, Bulletin of the Astronomical Society of India, 32, 335

\bibitem[{{Green}(2014)}]{green14a}
{Green}, D.~A. 2014, Bulletin of the Astronomical Society of India, 42, 47

\bibitem[{{Green}(2015)}]{green15}
{Green}, D.~A. 2015, \mnras, 454, 1517

\bibitem[{{Green}(2022)}]{green22}
{Green}, D.~A. 2022, {A Catalogue Galactic Supernova Remnants (2022 December version)}

\bibitem[{{Helfand} {et~al.}(2006){Helfand}, {Becker}, {White}, {Fallon}, \& {Tuttle}}]{helfand06}
{Helfand}, D.~J., {Becker}, R.~H., {White}, R.~L., {Fallon}, A., \& {Tuttle}, S. 2006, \aj, 131, 2525

\bibitem[{{Hurley-Walker} {et~al.}(2019){Hurley-Walker}, {Gaensler}, {Leahy}, {Filipovi{\'c}}, {Hancock}, {Franzen}, {Offringa}, {Callingham}, {Hindson}, {Wu}, {Bell}, {For}, {Johnston-Hollitt}, {Kapi{\'n}ska}, {Morgan}, {Murphy}, {McKinley}, {Procopio}, {Staveley-Smith}, {Wayth}, \& {Zheng}}]{hurley-walker19}
{Hurley-Walker}, N., {Gaensler}, B.~M., {Leahy}, D.~A., {et~al.} 2019, \pasa, 36, e048

\bibitem[{{Jonas} \& {MeerKAT Team}(2016)}]{Jonas2016}
{Jonas}, J. \& {MeerKAT Team}. 2016, in MeerKAT Science: On the Pathway to the SKA, 1

\bibitem[{{Joung} {et~al.}(2009){Joung}, {Mac Low}, \& {Bryan}}]{joung09}
{Joung}, M.~R., {Mac Low}, M.-M., \& {Bryan}, G.~L. 2009, \apj, 704, 137

\bibitem[{{Kaplan} {et~al.}(2002){Kaplan}, {Kulkarni}, {Frail}, \& {van Kerkwijk}}]{kaplan02}
{Kaplan}, D.~L., {Kulkarni}, S.~R., {Frail}, D.~A., \& {van Kerkwijk}, M.~H. 2002, \apj, 566, 378

\bibitem[{{Li} {et~al.}(1991){Li}, {Wheeler}, {Bash}, \& {Jefferys}}]{li91}
{Li}, Z., {Wheeler}, J.~C., {Bash}, F.~N., \& {Jefferys}, W.~H. 1991, \apj, 378, 93

\bibitem[{{M}auch {et~al.}(2020){M}auch, {C}otton, {C}ondon, \& et~al.}]{DEEP2}
{M}auch, T., {C}otton, W., {C}ondon, J., \& et~al. 2020, ApJ, 888, 61

\bibitem[{{Maxted} {et~al.}(2019){Maxted}, {Filipovi{\'c}}, {Hurley-Walker}, {Boji{\v{c}}i{\'c}}, {Rowell}, {Haberl}, {Ruiter}, {Seitenzahl}, {Panther}, {Wong}, {Braiding}, {Burton}, {P{\"u}hlhofer}, {Sano}, {Fukui}, {Sasaki}, {Tian}, {Su}, {Cui}, {Leahy}, \& {Hancock}}]{maxted19}
{Maxted}, N.~I., {Filipovi{\'c}}, M.~D., {Hurley-Walker}, N., {et~al.} 2019, \apj, 885, 129

\bibitem[{{Ostriker} {et~al.}(2010){Ostriker}, {McKee}, \& {Leroy}}]{ostriker10}
{Ostriker}, E.~C., {McKee}, C.~F., \& {Leroy}, A.~K. 2010, \apj, 721, 975

\bibitem[{{Ostriker} \& {Shetty}(2011)}]{ostriker11}
{Ostriker}, E.~C. \& {Shetty}, R. 2011, \apj, 731, 41

\bibitem[{{Padoan} {et~al.}(2016){Padoan}, {Pan}, {Haugb{\o}lle}, \& {Nordlund}}]{padoan16}
{Padoan}, P., {Pan}, L., {Haugb{\o}lle}, T., \& {Nordlund}, {\AA}. 2016, \apj, 822, 11

\bibitem[{{Pinheiro Gon{\c c}alves} {et~al.}(2011){Pinheiro Gon{\c c}alves}, {Noriega-Crespo}, {Paladini}, {Martin}, \& {Carey}}]{pinheiro11}
{Pinheiro Gon{\c c}alves}, D., {Noriega-Crespo}, A., {Paladini}, R., {Martin}, P.~G., \& {Carey}, S.~J. 2011, \aj, 142, 47

\bibitem[{{Ranasinghe} \& {Leahy}(2022)}]{ranasinghe22}
{Ranasinghe}, S. \& {Leahy}, D. 2022, arXiv e-prints, arXiv:2209.04570

\bibitem[{{Reach} {et~al.}(2006){Reach}, {Rho}, {Tappe}, {Pannuti}, {Brogan}, {Churchwell}, {Meade}, {Babler}, {Indebetouw}, \& {Whitney}}]{reach06}
{Reach}, W.~T., {Rho}, J., {Tappe}, A., {et~al.} 2006, \aj, 131, 1479

\bibitem[{{Sidorin} {et~al.}(2014){Sidorin}, {Douglas}, {Palou{\v{s}}}, {W{\"u}nsch}, \& {Ehlerov{\'a}}}]{sidorin14}
{Sidorin}, V., {Douglas}, K.~A., {Palou{\v{s}}}, J., {W{\"u}nsch}, R., \& {Ehlerov{\'a}}, S. 2014, \aap, 565, A6

\bibitem[{{Stil} {et~al.}(2006){Stil}, {Taylor}, {Dickey}, {Kavars}, {Martin}, {Rothwell}, {Boothroyd}, {Lockman}, \& {McClure-Griffiths}}]{stil06}
{Stil}, J.~M., {Taylor}, A.~R., {Dickey}, J.~M., {et~al.} 2006, \aj, 132, 1158

\bibitem[{{Supan} {et~al.}(2018){Supan}, {Castelletti}, {Peters}, \& {Kassim}}]{supan18}
{Supan}, L., {Castelletti}, G., {Peters}, W.~M., \& {Kassim}, N.~E. 2018, \aap, 616, A98

\bibitem[{{Sushch} {et~al.}(2017){Sushch}, {Oya}, {Schwanke}, {Johnston}, \& {Dalton}}]{sushch17}
{Sushch}, I., {Oya}, I., {Schwanke}, U., {Johnston}, S., \& {Dalton}, M.~L. 2017, \aap, 605, A115

\bibitem[{{Tammann} {et~al.}(1994){Tammann}, {Loeffler}, \& {Schroeder}}]{tammann94}
{Tammann}, G.~A., {Loeffler}, W., \& {Schroeder}, A. 1994, \apjs, 92, 487

\bibitem[{{Trushkin}(2001)}]{trushkin01}
{Trushkin}, S.~A. 2001, in ESA Special Publication, Vol. 459, Exploring the Gamma-Ray Universe, ed. A.~{Gimenez}, V.~{Reglero}, \& C.~{Winkler}, 109--111

\bibitem[{{Voit}(1992)}]{voit92}
{Voit}, G.~M. 1992, \mnras, 258, 841

\bibitem[{{Weiler} \& {Sramek}(1988)}]{weiler88}
{Weiler}, K.~W. \& {Sramek}, R.~A. 1988, \araa, 26, 295

\bibitem[{{Whiteoak} \& {Green}(1996)}]{whiteoak96}
{Whiteoak}, J.~B.~Z. \& {Green}, A.~J. 1996, \aaps, 118, 329

\bibitem[{{Wright} {et~al.}(2010){Wright}, {Eisenhardt}, {Mainzer}, {Ressler}, {Cutri}, {Jarrett}, {Kirkpatrick}, {Padgett}, {McMillan}, {Skrutskie}, {Stanford}, {Cohen}, {Walker}, {Mather}, {Leisawitz}, {Gautier}, {McLean}, {Benford}, {Lonsdale}, {Blain}, {Mendez}, {Irace}, {Duval}, {Liu}, {Royer}, {Heinrichsen}, {Howard}, {Shannon}, {Kendall}, {Walsh}, {Larsen}, {Cardon}, {Schick}, {Schwalm}, {Abid}, {Fabinsky}, {Naes}, \& {Tsai}}]{wright10}
{Wright}, E.~L., {Eisenhardt}, P.~R.~M., {Mainzer}, A.~K., {et~al.} 2010, \aj, 140, 1868

\end{thebibliography}

\end{document}